\documentclass[twocolumn]{aastex63}

\usepackage{xcolor}
\usepackage{amsmath}
\usepackage{affils}

\newcommand{\blt}[2]{\textbf{#1:} #2}
\newcommand{\reff}[0]{${\rm R_{eff}}$}
\newcommand{\sersic}[0]{S\'ersic}

\newcommand{\thh}[0]{$^{\rm th}$}

\newcommand{\fcr}[1]{\textcolor{black}{#1}}
\newcommand{\rrr}[1]{\textcolor{black}{#1}}
\newcommand{\rrrtwo}[1]{\textcolor{black}{#1}}
\newcommand{\rrrthree}[1]{\textcolor{black}{#1}}
\newcommand{\ndd}[0]{non-disky}
\newcommand{\NDD}[0]{Non-Disky}

\begin{document}
\shortauthors{Kado-Fong et al.}

\title{The In-situ Origins of Dwarf Stellar Outskirts in FIRE-2}

\DeclareAffil{princeton}{Department of Astrophysical Sciences, Princeton University,Princeton, NJ 08544, USA}

\author[ 0000-0002-0332-177X]{Erin Kado-Fong}
\affiliation{Department of Astrophysical Sciences, Princeton University, 4 Ivy Ln, Princeton, NJ 08544, USA}
\author[0000-0003-3939-3297]{Robyn E. Sanderson}
\affiliation{Department of Physics and Astronomy, University of Pennsylvania, 209 S 33rd St, Philadelphia, PA 19104, USA}
\affiliation{Center for Computational Astrophysics, Flatiron Institute, 162 5th Ave., New York, NY 10010, USA}
\author[0000-0002-5612-3427]{Jenny E. Greene}
\affiliation{Department of Astrophysical Sciences, Princeton University, 4 Ivy Ln, Princeton, NJ 08544, USA}
\author[0000-0002-6993-0826]{Emily C. Cunningham}
\affiliation{Center for Computational Astrophysics, Flatiron Institute, 162 5th Ave., New York, NY 10010, USA}
\author{Coral Wheeler}
\affiliation{Department of Physics \& Astronomy, Cal Poly Pomona University, 3801 W. Temple Ave., Pomona, CA 91768, USA}
\author[0000-0003-2544-054X]{T.K. Chan}
\affiliation{Institute for Computational Cosmology, Department of Physics, Durham University, South Road, Durham DH1 3LE, UK}
\author[0000-0002-6871-1752]{Kareem El-Badry}
\affiliation{Department of Astronomy and Theoretical Astrophysics Center, University of California Berkeley, Berkeley, CA 94720, USA}
\author[0000-0003-3729-1684]{Philip F. Hopkins}
\affiliation{TAPIR, California Institute of Technology, Mailcode 350-17, Pasadena, CA 91125, USA}
\author[0000-0003-0603-8942]{Andrew Wetzel}
\affiliation{Department of Physics \& Astronomy, University of California, Davis, One Shields Ave., Davis, CA 95616, USA}
\author[0000-0002-9604-343X]{Michael Boylan-Kolchin}
\affiliation{Department of Astronomy, The University of Texas at Austin, 2515 Speedway, Stop C1400, Austin, Texas 78712-1205, USA}
\author{Claude-Andr\'{e} Faucher-Gigu\`{e}re}
\affiliation{Department of Physics and Astronomy, Northwestern University, 2145 Sheridan Road, Evanston, IL 60647, USA}
\author[0000-0003-1385-7591]{Song Huang}
\affiliation{Department of Astrophysical Sciences, Princeton University, 4 Ivy Ln,  Princeton, NJ 08544, USA}
\author{Eliot Quataert}
\affiliation{Department of Astrophysical Sciences, Princeton University, 4 Ivy Ln,  Princeton, NJ 08544, USA}
\author[0000-0003-2539-8206]{Tjitske Starkenburg}
\affiliation{Department of Physics and Astronomy, Northwestern University, 2145 Sheridan Road, Evanston, IL 60647, USA}

\correspondingauthor{Erin Kado-Fong} 
\email{kadofong@princeton.edu}
  
  \date{\today}

{}

\begin{abstract}
Extended, old, and round stellar halos appear to be ubiquitous 
around high-mass dwarf galaxies ($10^{8.5}<M_\star/M_\odot<10^{9.6}$) 
in the observed universe. However, it is unlikely that these dwarfs
have undergone a sufficient number of minor mergers to form stellar
halos that are composed of predominantly accreted stars. 
Here, we demonstrate
that FIRE-2 (Feedback in Realistic Environments) cosmological zoom-in
simulations are capable of producing dwarf galaxies with realistic structure, including both a thick disk and round stellar halo.
Crucially, these stellar halos are formed in-situ, 
largely via the outward migration of disk stars. However, there 
also exists a large population of ``\ndd{}'' dwarfs in FIRE\rrrthree{-2} that lack
a well-defined disk/halo and do not resemble the observed dwarf population. 
These \ndd{} dwarfs tend to be either more gas poor or to have
burstier recent star formation histories than the disky dwarfs, suggesting
that star formation feedback may be preventing disk formation. 
Both classes of dwarfs underscore the power of a galaxy's intrinsic shape -- which is a direct quantification of the distribution of the 
galaxy's stellar content -- to interrogate the feedback implementation
in simulated galaxies.
\end{abstract}

\section{Introduction}
It has long been observed that dwarf galaxies host extended,
round old stellar populations reminiscent in structure to the stellar
halos of massive galaxies \citep{lin1983,minniti1996,minniti1999,zaritsky2000,aparicio2000, aparicio2000a,hidalgo2003, demers2006, bernard2007, stinson2009, strader2012, nidever2019, nidever2019a, pucha2019, kadofong2020}. 
Recently, it has been shown that high-mass dwarfs ($10^{8.5}<M_\star<10^{9.6} M_\odot$) can form a thick stellar \& HI disk 
\citep{roychowdhury2013, vanderwel2014, nathpatra2020} 
in conjunction with a round stellar halo \citep{kadofong2020}. 

In massive \fcr{($\rm M_\star>10^{10}M_\odot$)} galaxies, stellar halos are thought to form \fcr{primarily} from the accretion of
less massive satellite galaxies \citep[see, e.g.][]{bullock2005}.
However, dwarf galaxies do not \fcr{typically}
accrete enough stellar mass via minor mergers to form a stellar halo
in the same way as their massive counterparts \citep{read2006, purcell2007, brook2014, anglesalcazar2017, fitts2018}, and major
mergers are too rare to form dwarf stellar halos that are observable in
wide-field imaging \citep{kadofong2020}.
\fcr{Observed dwarf stellar halos also differ from massive stellar halos in that they} host intermediate-age stellar 
populations \citep[see, e.g., ][]{zaritsky2000,aparicio2000,aparicio2000a,hidalgo2003, demers2006,bernard2007,stinson2009,strader2012,nidever2019a,nidever2019,pucha2019}. These
observational results
suggest that in-situ processes might be responsible for the apparent ubiquity of their round and old stellar outskirts. 
Mechanisms that have been proposed include the star formation-driven
radial migration of stars \citep{stinson2009, elbadry2016}, and dynamical heating 
from interactions with dark subhalos \citep{starkenburg2015, starkenburg2016}. 
Due to their shallow potential wells (relative to
more massive galaxies), energy and momentum injected into the interstellar medium (ISM)
from star formation feedback can drive galactic outflows that displace a significant 
fraction of gas to large radii, thus causing fluctuations in the 
gravitational potential. The effect on the distribution of dark matter by
these baryon-driven fluctuations has been often cited 
as a potential resolution for the ``core-cusp'' problem 
\citep[see, e.g.][]{navarro1996, read2005, pontzen2012, chan2015}, but the same effect has also been shown
to induce stellar migration in previous studies. Notably, \cite{stinson2009} named
stellar radial migration as a potential formation pathway for stellar halos
in dwarfs, and \cite{elbadry2016} showed, using FIRE-1 simulations, that this radial migration operates 
most efficiently for dwarfs of stellar mass $M_\star \sim 10^{9}M_\odot$.
It is thus reasonable to say that a plausible formation pathway exists for 
in-situ stellar halo assembly at low masses \fcr{($M\star \lesssim 10^9 M_\odot$)}, and certainly that a mechanism
to \fcr{perturb} star particles to large radii operates in the FIRE\rrrthree{-2} simulations.

Observational studies also indicate that dwarf stellar halos may be \fcr{primarily} in-situ structures. First, observations of resolved stellar populations in the extended stellar outskirts of many local dwarfs galaxies show a population of both old and intermediate
age stars \citep[see ][and references therein]{stinson2009, vansevicius2004, hargis2020}, in
significant contrast to the ancient stellar halos of higher mass galaxies. This
indicates that the movement of stars into the extended envelope is an ongoing
process. Moreover, \citet{kadofong2020} show that, for a sample of 5974 dwarf galaxies at $z\lesssim0.15$, low-mass galaxies ($10^{8.5}<M_\star<10^{9.6} M_\odot$)
are generically rounder \rrr{at larger radius} regardless
of environment. They further show that \fcr{the majority of} isolated dwarfs host thick disks near their 
centers, while red satellite dwarfs are more spheroidal (following the familiar
color-morphology bimodality seen at higher masses).
\fcr{Because dwarfs do not typically accrete sufficient stellar mass to form an ex-situ stellar halo that would be detectable in current-generation widefield imaging \citep[e.g.][]{purcell2007},} 
both the dominance of this disk to halo structure in the isolated
dwarf sample and the presence of spheroidal stellar outskirts regardless of environment suggest
that the formation of low-mass stellar outskirts is a \fcr{primarily} in-situ process.
Thus, in order to match observations, simulated dwarfs should have 
both a young stellar disk and an 
old, round stellar halo. This requirement is especially relevant 
for a stellar halo produced largely by star formation feedback, as under-regulated or
over-active star formation feedback has been shown to be capable of disrupting the
disk in dwarf galaxies \citep{elbadry2018bradford, elbadry2018quataert, smith2020}. 
The requirement to match galaxy structure places a new
constraint on the
feedback physics implemented in simulations -- there
must be sufficient energy to heat the old stellar population, producing 
a round stellar halo, while simultaneously maintaining a young stellar disk. 

 The FIRE\footnote{\url{http://fire.northwestern.edu}} project combines both the resolution needed to study the detailed structure of
dwarf galaxies and the cosmological context necessary to understand the 
formation of that structure. 
In this work, we examine the three-dimensional stellar structure of a set of \fcr{isolated} dwarf galaxies 
($10^8\lesssim M_\star/M_\odot\lesssim10^{10}$) in the FIRE-2 simulation suite \citep{hopkins2018_main}.
We first determine whether or not the FIRE\rrrthree{-2} simulations reproduce
the disk-halo systems observed to be the dominant population in observations, and
then we examine the origin of these stellar halos. In \autoref{s:methods}, we 
summarize the basic properties of the FIRE\rrrthree{-2} simulations (\autoref{s:methods:fire}) 
and our intrinsic shape measurement method (\autoref{s:methods:intrinsicshape}).
We present our main findings in \autoref{s:results}, then discuss those results in 
\autoref{s:discussion}. In particular, a comparison of this study to previous theoretical
works can be found in \autoref{s:discussion:comparison}, while a discussion of the
origin of stellar halos in the FIRE\rrrthree{-2} dwarfs and the context of non-disky dwarfs is found in
\autoref{s:discussion:formation} and \autoref{s:discussion:triaxial}, respectively. 

\begin{figure*}[htb]
\centering     
\includegraphics[width=.8\linewidth]{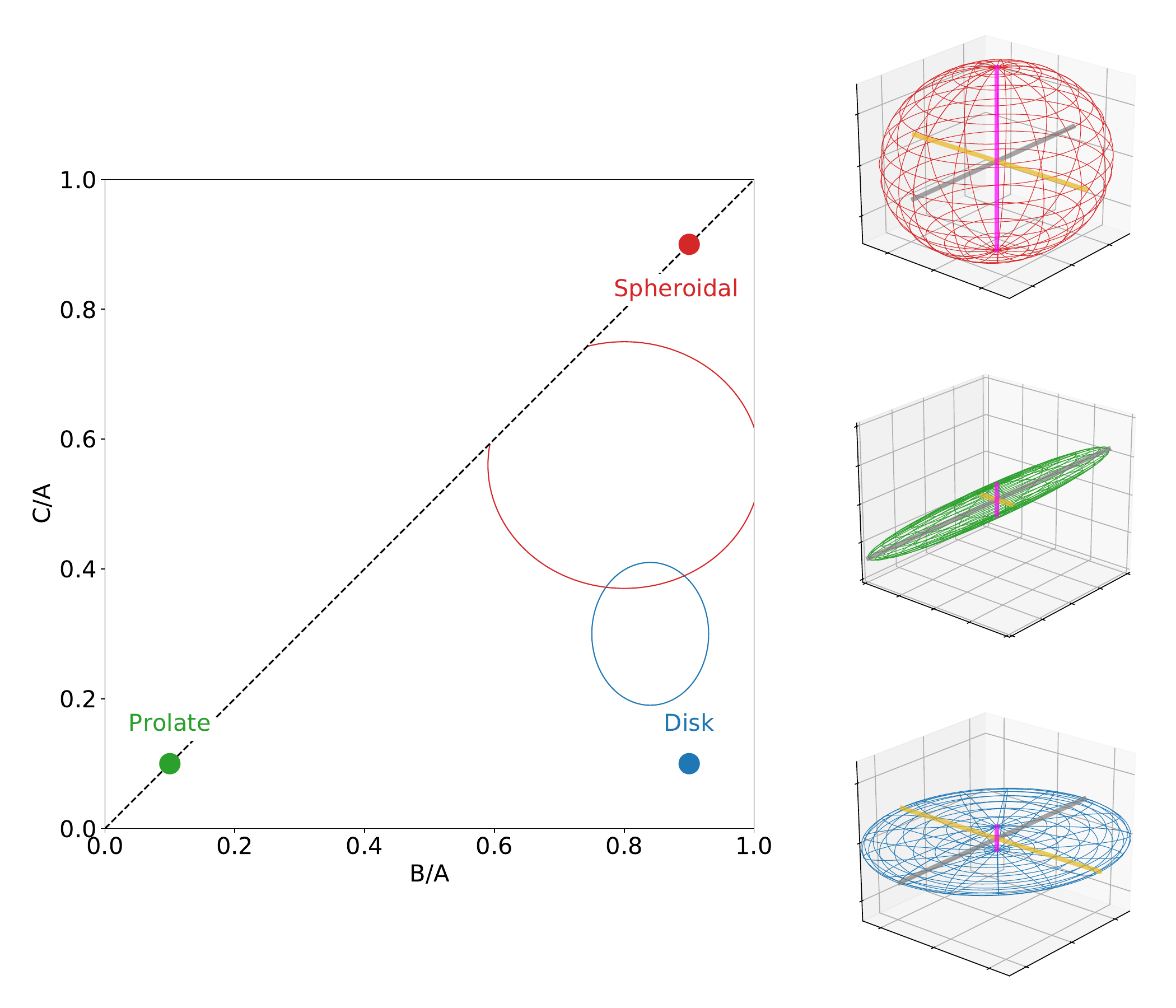}
\caption{
    A schematic diagram to illustrate movement in the $B/A$ vs. $C/A$
    plane. The red, green, and blue points show the position of
    an archetypal spheroid, prolate ellipsoid, and disk, respectively. The
    axis ratios are ($B/A$,$C/A$) = (0.9,0.9), (0.1,0.1), and
    (0.9, 0.1) for the three cases.  The observed 1$\sigma$ distribution of the dwarf sample in
    \cite{kadofong2020} as measured at 1$\rm R_{eff}$ and 4$\rm R_{eff}$
    are shown by blue and red ellipses,
    respectively.
    At right, we show a three-dimensional
    representation of the ellipsoid that corresponds to each case 
    in the corresponding color. We additionally show the principal 
    axes A, B, and C as grey, gold, and magenta lines in each panel.
    }
\label{f:schematic}
\end{figure*}

\begin{figure*}[htb]
\centering     
\includegraphics[width=\linewidth]{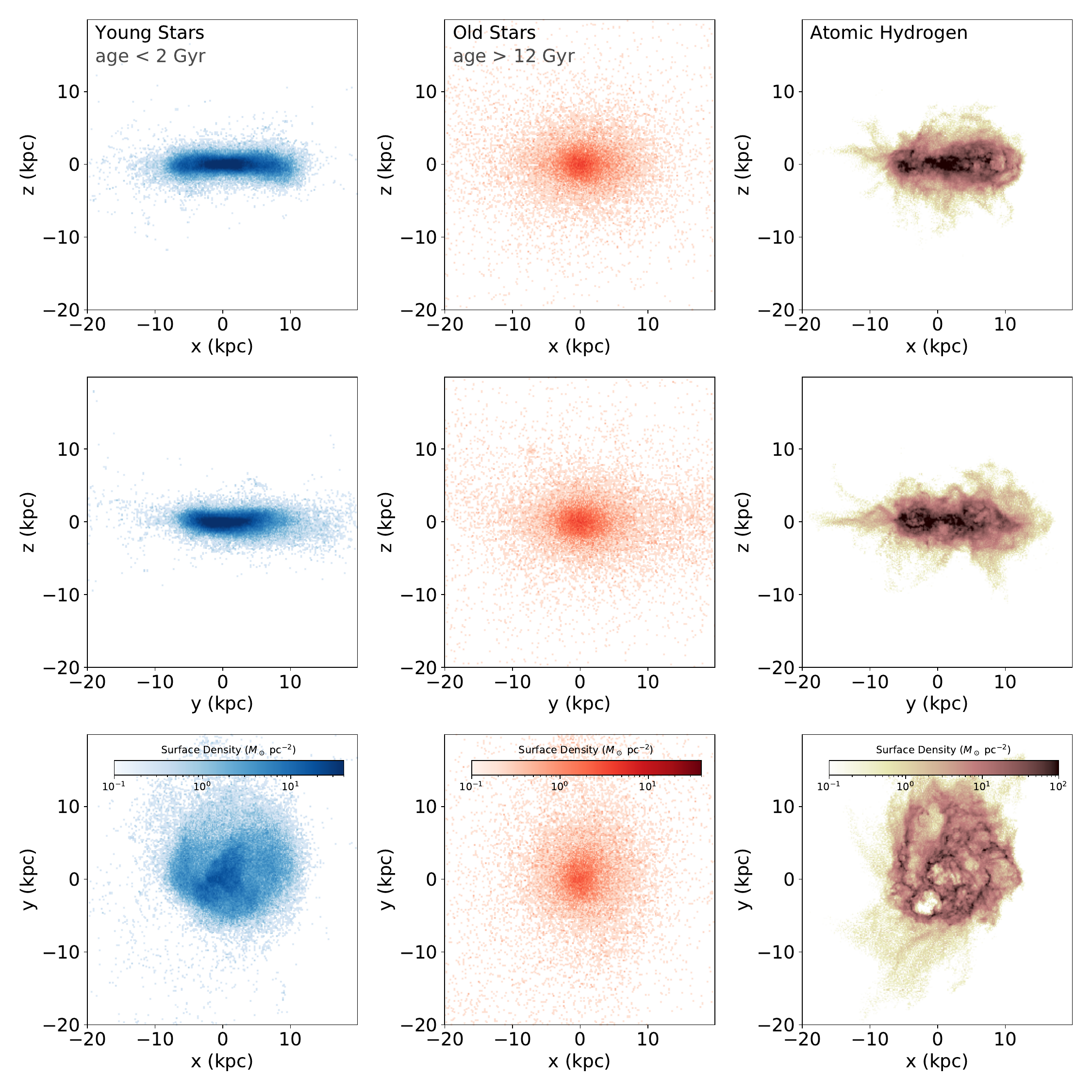}
\caption{
    The surface density of young stars (left column, star particles with ages less than
    2 Gyr), old stars (middle column, star particles with ages greater than 12 Gyr), and
    HI (right column) for an example dwarf with a disk-halo structure, m11h (MHD+). 
    The rows show the projection along the y-axis, x-axis, and z-axis,
    \fcr{such that the z-axis is aligned with the disk minor axis}.
    }
\label{f:projections_m11h}
\end{figure*}

\begin{figure*}[htb]
\centering     
\includegraphics[width=\linewidth]{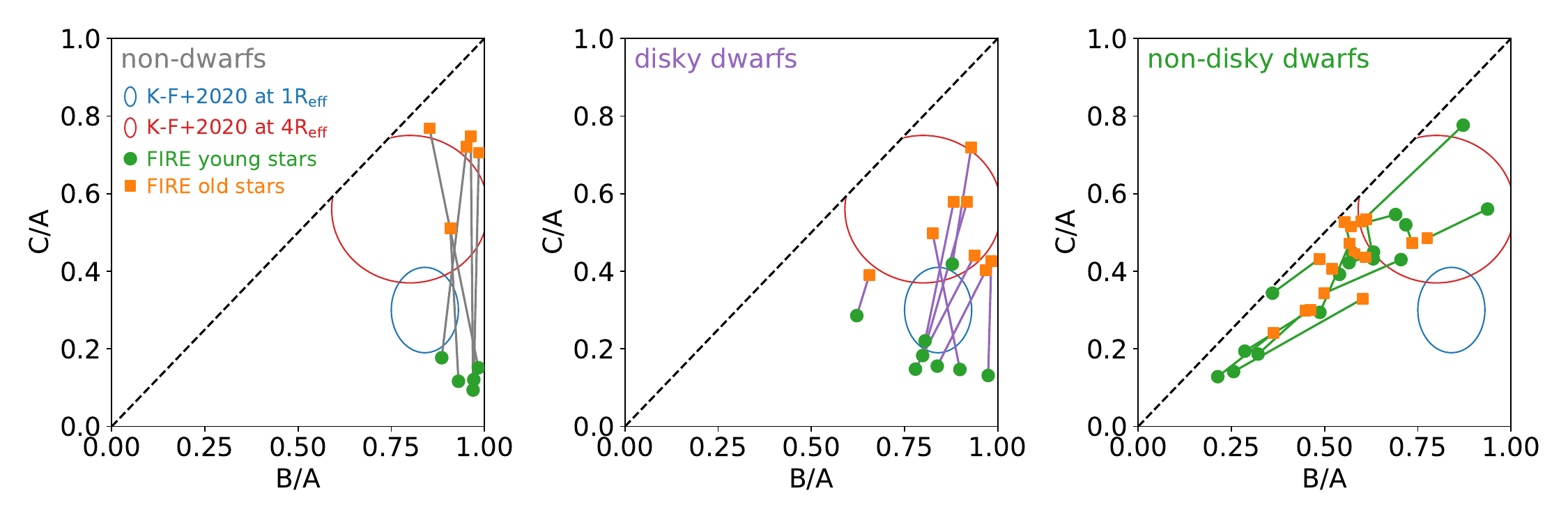}
\caption{
    The intrinsic shape of the young and old stellar populations for the three subsets
    of the sample we consider in this work. In each panel, the   green filled circles show the 
    intrinsic shape of the young \fcr{(ages less than 2 Gyr)}
    stellar population and the orange squares show the intrinsic
    shape of the old stellar population \fcr{(ages older than 12 Gyr)}. 
    Lines connect the measurements for each individual
    galaxy.
    The blue and red ellipses show the observed
    1$\sigma$ distribution of the dwarf sample of \cite{kadofong2020} at 1\reff{}
    and 4\reff{}, respectively.
     At left a comparison sample of high-mass ($\rm M_\star>10^{10}M_\odot$) galaxies
    is shown; these galaxies host a clear disk-halo transition between their young and
    old stellar populations, where the youngest stars are assembled in a thin disk and the
    oldest populate a spheroidal stellar halo. The center panel shows the sample of dwarf 
    galaxies that host a disk halo system; the disks in these dwarfs are thicker (higher C/A)
    than those of the higher mass galaxies, while the old stellar components are less round 
    (lower C/A). Finally, at right, we show the \ndd{} dwarfs in the sample, \fcr{which are characterized by the lack of a young
    stellar disk, as well as
    the lack of a monotonic increase in C/A and the
    presence as a significant change in B/A
    function of age}. 
    }
\label{f:baca}
\end{figure*}

\section{Methods}\label{s:methods}
Here we give a brief overview of the FIRE\rrrthree{-2} simulation suite and the sample of FIRE\rrrthree{-2} galaxies
included in this work. We then detail the method used to compute the intrinsic shape of
the simulations.

\subsection{The FIRE Simulations}\label{s:methods:fire}
Due to the low masses and small sizes of dwarf galaxies, studying their detailed structure
requires high resolution simulations. Moreover,
their shallow potential wells make the structure of 
dwarf galaxies relatively more sensitive to the details of the implementation of feedback from
star formation, such as stellar winds and supernovae, than the structure of more massive galaxies. Thus, while
realistic feedback prescriptions are crucial to understanding
the structural properties of low mass galaxies, such galaxies are also among the most sensitive tests
of these same feedback prescriptions \citep[e.g.][]{brooks2014, hu2016, wheeler2017, hu2019, wheeler2019, dashyan2020, smith2020}. Furthermore, \emph{cosmological} simulations are \rrr{necessary to} properly capture the significant effect of reionization on these small galaxies \citep[e.g.][]{2009ApJ...693.1859B,2015MNRAS.453.1503B,2017MNRAS.469L..83W,2017MNRAS.471.3547F,2019MNRAS.490.1186G}.
\rrr{Cosmological simulations also supply} the variety of environments \citep[e.g.][]{2019MNRAS.489.4574G,2019MNRAS.489.5348J} and assembly histories \citep[e.g.][]{2017MNRAS.471.3547F} \rrr{needed to 
study the dwarf population}, 
especially given the relatively large impact that interactions \rrr{even}
with purely dark subhalos can have on the star formation histories of dwarfs \citep{starkenburg2016}.

We use simulations from the FIRE project, specifically the suite of
cosmological-baryonic zoom-in simulations run with the FIRE-2 \rrrthree{feedback models} \citep{hopkins2018_main} along with some variations described below. \rrrthree{The FIRE project contains a
suite of simulated dwarf galaxies that are uniquely well-suited for
the structural study at hand -- as we have argued above,
assessing the origin of dwarf stellar
structure formation requires simulations that are both cosmological
in nature and high resolution. The baryonic particle mass of
the dwarf simulations at hand range between 2100-7100 M$_\odot$, with a
minimum gravitational force softening of $\sim 2$ pc. Though there are
a relatively low number of initial conditions (we study nine in this work),
the resolution of the FIRE-2 simulations is critical for robust
studies of the stellar structure of low-mass galaxies \citep{ludlow2021}. 
Indeed, it has been shown that the stellar structure of FIRE-2 dwarfs at even
lower masses remain robust to resolution effects \citep{wheeler2019}.
Thus, we expect the structure of these relatively more massive
galaxies to be robust to resolution effects due to the high resolution
of the FIRE-2 simulations -- moreover, as a small forward reference, we
note that we find no evidence for a dependence between resolution and
the dwarf properties we measure in this work.}

All simulations use the Meshless-Finite-Mass (hereafter MFM) mode of the GIZMO\footnote{\url{http://www.tapir.caltech.edu/~phopkins/Site/GIZMO.html}} gravity+magnetohydrodynamic
code \citep{hopkins2015}, which provides adaptive spatial resolution, conservation of mass, energy, and momentum, and excellent shock-capturing and conservation of angular momentum, reproducing advantages of both smoothed-particle hydrodynamics (SPH) and Eulerian adaptive mesh refinement (AMR) schemes. Gravity is solved with an improved version of the Tree-PM solver from GADGET-3 \citep{Springel2005gadget2}, with fully-adaptive (and fully-conservative) gravitational force softenings for gas (so hydrodynamic and force softenings are always self-consistently matched), following \citet{Price2007}.

Table \ref{t:basic} lists the simulations included in this study. All are simulations of isolated galaxies (the main halo being simulated is the highest-mass object in the zoomed box). The names of the simulations follow a convention: (1) in the prefix ``mXX,'' the number XX denotes the order of magnitude of the halo mass of the most massive halo in the zoomed region, in solar masses; (2) the letter following the prefix is a label representing the simulation's unique initial conditions; (3) the parenthetical after the resolution indicator is
a shorthand reference to the physics included in
the run (described further below).

We select \fcr{the $z=0$ snapshots} of the set of isolated dwarf galaxy simulations with stellar
masses $10^8 \lesssim M_\star/M_\odot < 10^{10}$, along with a set of 
three higher mass galaxies ($M_\star>10^{10}M_\odot$) that we use as
a reference sample.
We will show in this work that the stellar halos of the high-mass 
reference set are built largely from accreted particles, as expected for $\rm L_\star$ galaxies,
and use this high-mass sample as a benchmark for classical stellar halo assembly.

We use simulations run with three different variations of the FIRE physics engine, denoted in the ``Run Type'' column of Table \ref{t:basic} as (Hydro+, no MD), (Hydro+, MD), (MHD+),
and (CR+) to explore the sensitivity of our conclusions to small changes in the baryonic physics implemented in the simulations. Runs marked (Hydro+, no MD) and (Hydro+, MD) use the core FIRE\rrrthree{-2} physics described in full in \citet{hopkins2018_main} with and without
metal diffusion (MD)\rrrthree{, respectively.}
Runs marked (Hydro+, MD) in Table \ref{t:basic} use the physics described above, but also include the numerical implementation of turbulent metal diffusion described in \citet{escala}. \rrrthree{Including a model for turbulent metal diffusion allows neighboring gas cells to exchange gas as it is enriched, as we presume happens in nature. This has the main effect of narrowing the distribution of abundances of different metals in the gas and stars, bringing the width of the abundance distributions in line with observations from dwarf galaxies without changing the structural properties of the galaxies themselves \citep{escala}.}
\fcr{\rrrthree{Since w}e do not expect metal diffusion to affect the properties \rrrthree{studied} in this work, both (Hydro+) designations may be treated interchangeably\rrrthree{, as separate realizations of similar dwarfs from the same initial conditions that differ only in the stochastic variation induced by a different pattern of stellar feedback---}we propagate the presence of metal diffusion in the naming scheme for completeness.} 

In brief, radiative heating and cooling is treated from $10-10^{10}\,$K, including free-free, photo-ionization/recombination, Compton, photoelectric \& dust collisional, cosmic ray, molecular, and metal-line \& fine-structure processes (following each of 11 tracked species independently), and accounting for photo-heating both by a UV background \citep{FaucherGiguere2009} and local sources, as well as self-shielding. 
\fcr{In the UV background model used for the simulations \rrrthree{ in this work}, reionization occurs at $z\sim10$, significantly ear\rrrthree{l}ier than current
empirical constraints \citep[see, e.g.][]{planck2020}.}\rrrthree{ The early reionization onset mainly affects the star formation histories of ultra-faint galaxies (up to around $M_* \sim 10^5\ M_{\odot}$; \citealt{wheeler2019}) with far less stellar mass than the simulated galaxies studied here.}

Star formation occurs only in gas identified as self-gravitating according to the \citet{Hopkins2013sf_criteria} criterion, which is also molecular and self-shielding (following \citealt{Krumholz2011}), Jeans unstable, and exceeds a minimum density threshold $n_{\rm min}=1000\,{\rm cm^{-3}}$. Once a star particle forms, the simulations explicitly follow several different stellar feedback mechanisms, including (1) local and long-range momentum flux from radiation pressure (in the initial UV/optical single-scattering, and re-radiated light in the IR), (2) energy, momentum, mass and metal injection from SNe (Types Ia and II) and stellar mass loss (both OB and AGB), and (3) photo-ionization and photo-electric heating. Every star particle is treated as a single stellar population with known mass, age, and metallicity from which all feedback event rates, luminosities, energies, mass-loss rates, and other relevant quantities are tabulated directly from stellar evolution models ({\sc starburst99}; \citealt{leitherer1999}), assuming a \citet{Kroupa2001} IMF.

 Runs marked (MHD+) again use the same physics as the (Hydro+) runs but solve the equations of ideal magneto-hydrodynamics (MHD) as described and tested in \citet{hr2016} and \citet{hopkins2016}, with anisotropic Spitzer-Braginskii conduction and viscosity as described in \citet{hopkins2017}, \citet{su2017}, and \citet{hopkins2020}. Runs marked (CR+) include all of the physics
 implemented in the (MHD+) runs, with the addition of the magnetohydrodynamic treatment of cosmic rays described in 
 \citet{chan2019}, \citet{hopkins2020}, and \citet{ji2020}. 
 \rrrthree{Magnetic fields, conduction, and viscosity have only 
 small effects on bulk galaxy properties at any mass scale 
 \citep{hopkins2020}, but cosmic rays can can suppress SF and 
 total stellar masses by factors $\sim 2-4$ for galaxies with 
 $M_\textrm{vir} \gtrsim 10^{11} M_{\odot}$ at late times ($z 
 \lesssim 1-2$) by building up hot gas in the galaxy halo that 
 suppresses the galactic fountain \citep{hopkins2020}. This only 
 occurs in halos massive enough to confine the hot ($T\sim10^6 
 K$) gas, and the set of dwarfs we consider here are right on the 
 boundary of this effect, which is more pronounced in the more 
 massive systems used for comparison (last five rows of Table 
 \ref{t:basic}. However, we do not see any systematic difference 
 in the shapes of the galaxies produced in the (CR+) runs
 relative to the (MHD+) and (Hydro+) runs.}
 
These simulations are uniquely well-suited to study the connection between the formation of dwarf galaxies and their structure. The combination of physics implementations in the (Hydro+) runs have already been shown to reproduce the mass-size relation of observed galaxies across $\gtrsim5$ orders of magnitude in stellar mass \citep{elbadry2016,chan2018}. The sample includes a large range of different assembly histories (9 in total for the dwarf sample) that give rise to a wide variety of present-day galaxy morphologies, from thin disks to highly diffuse structures \citep{elbadry2018}.
\fcr{\rrrthree{With the FIRE-2 feedback model}, the resolved ISM produces star formation histories that are generically bursty for dwarf galaxies \citep[e.g.][]{sparre2017, fauchergiguere2018,florevelazquez2021}, an effect which can be suppressed in lower-resolution simulations that \rrrthree{do not resolve the multiphase ISM.} On the other hand, the burstiness of the star formation in these simulations is slightly enhanced relative to models that explicitly include local radiative feedback, which has the effect of warming the gas and disrupting some GMCs before collapse, leading to a smoother SFH in massive dwarfs \citep{2020MNRAS.491.3702H}.}

\rrrthree{Because these simulations resolve individual large GMCs \citep{2020MNRAS.497.3993B} where the star formation occurs, the star formation has a very high spatial granularity across the galaxy; clustering of starforming clouds is thus naturally taken into account down to $\sim$pc scales. The granularity of the resulting stellar feedback in both space and time is also very high, since the }mass of the individual star particles \rrrthree{(2100-7100 $M_\odot$ per particle)} is sufficiently small to support the assumption that each represents a single-age, single-metallicity stellar population while still being large enough to fully sample the IMF at the high-mass end \citep{sanderson2020}. The inclusion of turbulent metal diffusion additionally reproduces the observed dependence of the width of abundance spreads on stellar mass in dwarfs \citep{escala}. Comparing runs with additional physics to the (Hydro+) runs for identical initial conditions allows us to both confirm that these variations do not strongly influence the structure of dwarfs, and set a bound on the degree of scatter due to stochastic supernova feedback for a fixed assembly history.

\defcitealias{chan2018}{1}
\defcitealias{hopkins2020}{2}
\defcitealias{hopkins2018_main}{3}
\defcitealias{elbadry2018}{4}
\defcitealias{wetzel2016}{5}

\begin{deluxetable*}{lccccccc}
\tablecaption{Basic Properties of the FIRE-2 Galaxies Included in this Work}
\tablehead{
\colhead{Run Name}&
\colhead{$\rm M_{\rm vir}$} &
\colhead{$\rm R_{vir}$} &
\colhead{$\rm M_{\rm \star, 90}$} &
\colhead{$\rm R_{50,\star}$} &
\colhead{Morphology} &
\colhead{Baryonic particle mass} &
\colhead{Reference} \\
\colhead{} &
\colhead{[$\rm 10^{11} M_\odot$]} &
\colhead{ [kpc] } &
\colhead{ [$\rm 10^{9} M_\odot$]} &
\colhead{ [kpc]} &
\colhead{} &
\colhead{[$\rm 10^{3} M_\odot$]} &
\colhead{} 
}
\startdata
m11a (Hydro+, no MD) &   0.39 &  0.89 &   0.11 &  2.79 &   \ndd{} &  2.1 &          \citepalias{chan2018} \\
m11a (CR+)           &   0.39 &  0.89 &   0.05 &  1.27 &   \ndd{} &  2.1 &       \citepalias{hopkins2020} \\
m11b (CR+)           &   0.40 &  0.90 &   0.09 &  2.20 &      disky &  2.1 &          \citepalias{chan2018} \\
m11b (MHD+)          &   0.40 &  0.90 &   0.07 &  2.29 &      disky &  2.1 &       \citepalias{hopkins2020} \\
m11b (Hydro+, no MD) &   0.40 &  0.90 &   0.11 &  2.54 &      disky &  2.1 &  \citepalias{hopkins2018_main} \\
m11c (CR+)           &   1.37 &  1.35 &   0.78 &  2.79 &   \ndd{} &  2.1 &       \citepalias{hopkins2020} \\
m11c (MHD+)          &   1.41 &  1.36 &   1.16 &  3.21 &   \ndd{} &  2.1 &       \citepalias{elbadry2018} \\
m11c (Hydro+, no MD) &   1.37 &  1.35 &   0.84 &  3.00 &   \ndd{} &  2.1 &       \citepalias{elbadry2018} \\
m11d (CR+)           &   2.71 &  1.70 &   1.55 &  4.12 &   \ndd{} &    7 &       \citepalias{hopkins2020} \\
m11d (MHD+)          &   2.75 &  1.71 &   5.03 &  5.71 &   \ndd{} &    7 &       \citepalias{hopkins2020} \\
m11d (Hydro+, MD)    &   2.72 &  1.70 &   4.06 &  6.97 &   \ndd{} &  7.1 &       \citepalias{hopkins2020} \\
m11e (Hydro+, MD)    &   1.43 &  1.38 &   1.46 &  3.84 &   \ndd{} &  7.1 &       \citepalias{hopkins2020} \\
m11e (CR+)           &   1.40 &  1.36 &   0.65 &  3.28 &   \ndd{} &  7.1 &       \citepalias{elbadry2018} \\
m11e (MHD+)          &   1.45 &  1.38 &   1.25 &  4.55 &   \ndd{} &  7.1 &       \citepalias{elbadry2018} \\
m11h (CR+)           &   1.76 &  1.47 &   2.87 &  3.60 &      disky &    7 &        \citepalias{wetzel2016} \\
m11h (MHD+)          &   1.81 &  1.49 &   4.44 &  3.46 &      disky &    7 &       \citepalias{hopkins2020} \\
m11h (Hydro+, MD)    &   1.80 &  1.48 &   3.62 &  4.13 &      disky &  7.1 &       \citepalias{hopkins2020} \\
m11i (Hydro+, MD)    &   0.69 &  1.08 &   0.93 &  3.79 &   \ndd{} &  7.1 &       \citepalias{hopkins2020} \\
m11i (CR+)           &   0.63 &  1.05 &   0.22 &  2.88 &   \ndd{} &  7.1 &       \citepalias{hopkins2020} \\
m11i (MHD+)          &   0.68 &  1.07 &   0.58 &  3.68 &      disky &  7.1 &       \citepalias{hopkins2020} \\
m11q (MHD+)          &   1.45 &  1.38 &   1.87 &  3.04 &   \ndd{} &    7 &       \citepalias{hopkins2020} \\
m11q (Hydro+, MD)    &   1.41 &  1.36 &   0.63 &  2.62 &   \ndd{} &  7.1 &       \citepalias{hopkins2020} \\
m11v (CR+)           &   2.95 &  1.74 &   2.59 &  8.27 &   \ndd{} &  7.1 &       \citepalias{hopkins2020} \\
m11v (MHD+)          &   2.20 &  1.58 &   2.51 &  3.52 &      disky &  7.1 &       \citepalias{hopkins2020} \\
\hline
m11f (CR+)           &   4.30 &  1.98 &  12.02 &  3.56 &  non-dwarf &   12 &       \citepalias{hopkins2020} \\
m11f (MHD+)          &   4.75 &  2.04 &  31.74 &  2.57 &  non-dwarf &   12 &       \citepalias{hopkins2020} \\
m11g (CR+)           &   5.35 &  2.13 &  11.00 &  4.80 &  non-dwarf &   12 &  \citepalias{hopkins2018_main} \\
m11g (MHD+)          &   6.05 &  2.21 &  49.02 &  2.78 &  non-dwarf &   12 &       \citepalias{hopkins2020} \\
m12i (Hydro+, no MD) &  10.46 &  2.66 &  70.60 &  2.89 &  non-dwarf &  7.1 &       \citepalias{hopkins2020} \\
\enddata
\tablecomments{Virial mass ($\rm M_{Vir}$) is measured within the \citet{bn1998} virial radius of the halo at $z=0$. \rrr{$\rm M_{\star,90}$ and $\rm R_{\star,50}$ are the 
stellar mass and radius enclosing 90\% and 50\% of the stellar mass within
30 kpc of the galaxy center, respectively.}
``Baryonic particle mass'' denotes the initial mass of gas \& star particles in the simulation. ``Hydro+'' indicates runs with the 
core physics suite with and without metal diffusion (``MD''), 
while ``MHD+'' indicates runs that also
include treatment for magnetic fields and fully anisotropic conduction and viscosity. ``CR+''
indicates runs that also include treatment of cosmic rays (in addition to the physics described
by MHD+).
Morphological classifications are discussed in \autoref{s:results:intrinsicshapes}.
References: (1) \citealt{chan2018}, (2) \citealt{hopkins2020},
(3) \citealt{hopkins2018_main}, (4) \citealt{elbadry2018}, (5) \citealt{wetzel2016}. }
\label{t:basic}
\end{deluxetable*}

\subsection{Intrinsic Shape Measurements}\label{s:methods:intrinsicshape}
We parameterize the three dimensional shapes of the sample using the
ratio of the semi-principal axis diameters A, B, and C where
$\rm A \geq B \geq C$. The ratios of these axes, $\rm B/A$ and $\rm C/A$, give
a quantitative description of the intrinsic shape of each galaxy; in 
\autoref{f:schematic}, we show three extreme cases in both $\rm B/A$ vs.
$\rm C/A$ space at left and as wireframe renderings at right. In particular,
we show a prolate galaxy in green ($\rm C/A \sim B/A \ll 1$), a disky galaxy in
blue ($\rm C/A \ll B/A \sim 1$), and a spheroidal galaxy in red 
($\rm C/A \sim B/A \sim 1$). For context, we also show the $1\sigma$ contours of
the observed dwarfs sample of \cite{kadofong2020} in blue (measurements at
1\reff{}) and red (measurements at 4\reff{}).

Making structural measurements for age-separated stellar components
of low-mass halos requires simulations of sufficiently high resolution.
Even at the mass resolution of the FIRE\rrrthree{-2} simulations studied in this work,
we use cumulative \fcr{(i.e. the shape of the stars within a given radius)},  
rather than differential \fcr{(the shape of the stars at a given radius)}, 
intrinsic shape measurements
in order to overcome instability due to low stellar densities in the 
outskirts of the dwarfs, \rrrthree{though we note that
the radial density profile remains stable}. 
This is a qualitatively different approach
than what is used in integrated light measurements of intrinsic shape, 
where the galaxy intrinsic shape distribution is inferred at fixed 
radius \citep{padilla2008, vanderwel2014, kadofong2020}. \fcr{The need for a
cumulative measure of the intrinsic shape will also lead us to measure intrinsic
shapes as a function of stellar age instead of galactocentric radius, comparing in
particular the young (ages $<2$ Gyr) and old (ages $>12$ Gyr) stellar populations (see 
\autoref{s:results:intrinsicshapes}). Based on observations of nearby dwarfs,
this age separation is expected to 
correlate with the observational differential shapes
\citep[see, e.g.][]{zaritsky2000,aparicio2000,aparicio2000a,hidalgo2003, demers2006,bernard2007,stinson2009,strader2012,nidever2019a,nidever2019,pucha2019}.
However, we note that the age-separated populations that we use in this work are not the
exact equivalent to the radially-separated stellar populations of \cite{kadofong2020}.}
The simulated and observed measurements also differ in that the observed 
measurements \fcr{of the galaxy projected shapes 
are used to infer the intrinsic shape distribution of the galaxy
population} 
(whereas the simulated values are direct intrinsic shape measurements of individual
galaxies), 
and that the simulated measurements are mass-weighted
in bins of stellar age (whereas the observed measurements are light-weighted at
fixed radii). 
{}
Thus, \fcr{the emphasis of this work will be on the change in galaxy
structure within the simulations and observations as a function of galaxy properties, 
rather than a direct quantitative comparison between the observed and simulated
measurements. }
\rrrtwo{However, in \autoref{s:appendix:inference}, we also 
demonstrate that these cumulative
measurements well-trace mock measurements made using the observational 
inference machinery of \cite{kadofong2020}.}

As mentioned above, unlike in observations, in simulations we are able to directly
measure the intrinsic shape of the galaxies in our sample. To do so,
we compute the reduced moment of inertia tensor $\mathbf{I}$ 
from the set of star particles
at distances less than $R_{\max}=50$ kpc for the simulated galaxies
with $\rm M_{\rm halo} \sim 10^{11} M_\odot$ and 
$R_{\max}=70$ kpc for the galaxy with $\rm M_{\rm halo} \sim 10^{12} M_\odot$. 
The
reduced moment of inertia tensor $\mathbf{I}$ is given by

\begin{equation}\label{e:inertia}
\mathbf{I}=
\begin{bmatrix}
I_{xx}& I_{xy}& I_{xz} \\
I_{yx}& I_{yy}& I_{yz} \\
I_{zx}& I_{zy}& I_{zz} 
\end{bmatrix}
\end{equation}
where each element is computed from the star particles (of mass $m_k$) as
\begin{equation}\label{e:iij}
I_{ij} = \frac{\sum_{r_k < R_{\max}} m_kq_{i,k}q_{j,k} / r_k}{\sum_{r_n < R_{\max}} m_n},
\end{equation}
for $i,j\in(x,y,z)$ and where $\mathbf{r_k} = (q_x^k,q_y^k,q_z^k)$ is
the distance from the principal halo center. \rrr{The principle axes 
calculated from the reduced moment of inertia may overweight star particles
at large distances relative to the unreduced moment of inertia tensor; for
our sample, however,
we find a median fractional difference in the \rrrthree{principal} axis ratios of 5\%, 
and conclude that our choice of
moment of inertia tensor does not significantly affect our results.}
The eigenvalues $\lambda_1$, $\lambda_2$ and $\lambda_3$ of this matrix
are the inverse squares of the semi-principal axis diameters $A$, $B$, and $C$
of the Poinsot ellipsoid that corresponds to the moment of inertia tensor: 

\begin{equation}
\begin{split}
    \frac{B}{A} &= \sqrt{\frac{\lambda_1}{\lambda_2}}\\
    \frac{C}{A} &= \sqrt{\frac{\lambda_1}{\lambda_3}}.
\end{split}    
\end{equation}

\fcr{This method is, in principle, sensitive to the choice of
$R_{\max}$ and emphasizes the contribution of particles at large
$r$ (see \autoref{e:iij}). We test the effect of our choice of
intrinsic shape measurement method as follows.}
First, though we choose a larger $R_{\max}$ for the higher mass halo, 
we would derive the same shape parameters $B/A$ and $C/A$ to 
within $\Delta(B/A)\sim\Delta(C/A)\sim0.01$ at $R_{\max}=50$ kpc.
\fcr{Furthermore, although we adopt a fixed value of
$R_{\max}=50$ kpc for the dwarf galaxies
considered in this work, we find that varying 
$R_{\max}$ does not qualitatively affect our 
results down to $R_{\max}\sim10$ kpc. Finally, 
we fit two dimensional \sersic{}
profiles to projections of each galaxy along its principal
axes, and find
no evidence for a systematic offset between the \sersic{} fits and 
moment of inertia-derived axis ratios. 
Thus, although the moment of inertia method
potentially overemphasizes star particles at large distances, 
we find that our results are not impacted 
by this bias in practice.}

Observational works indicate that dwarfs in this mass range should host a central thick disk
\citep[see, e.g.][]{vanderwel2014}
with a round extended
stellar component \citep{kadofong2020} that is populated by
intermediate/old stars \citep[see][and citations therein]{stinson2009}. 
If a disk-halo system is indeed 
present in the FIRE\rrrthree{-2} dwarfs, we should be able to detect the same transition from
disk to halo when computing the intrinsic shape in bins of stellar age \rrr{-- we demonstrate that this transition
is indeed detectable in mock HSC data in 
\autoref{s:results:observability}}. We visualize
an example of such a divide in \autoref{f:projections_m11h}, which shows the 
projected density of the dwarf galaxy m11h (MHD+) in young stars (left column, blue)
old stars (middle column, red) and neutral hydrogen (right column). 

\section{Results}\label{s:results}

\begin{deluxetable*}{lllllll}
\tablecaption{Mean Intrinsic Shape Ratios of the FIRE-2 Dwarf Sample}
\tablehead{
\colhead{} &
\colhead{$\langle \frac{B}{A}\rangle_{\rm young}$} &
\colhead{$\langle \frac{C}{A}\rangle_{\rm young}$} &
\colhead{$\langle \rm N \rangle_{young}$} &
\colhead{$\langle \frac{B}{A}\rangle_{\rm old}$} &
\colhead{$\langle \frac{C}{A}\rangle_{\rm old}$} &
\colhead{$\langle \rm N \rangle_{old}$} 
}
\startdata
Non-dwarfs  &  $0.95\pm0.02$ &  $0.13\pm0.01$ & $9.6\times10^5$ &  $0.93\pm0.02$ &  $0.69\pm0.04$ & $9.2\times10^5$ \\
Disky Dwarfs   &  $0.82\pm0.03$ & $0.21\pm0.03$ & $6.3\times10^4$ & $0.89\pm0.04$ &  $0.50\pm0.04$ & $8.6\times10^4$ \\
\NDD{} Dwarfs & $0.55\pm0.05$ & $0.39\pm0.04$ & $9.9\times10^4$ & $0.56\pm0.02$ & $0.42\pm0.02$ & $9.6\times10^4$ \\
\enddata
\tablecomments{Mean intrinsic shape measurements (denoted by $\langle\rangle$) and 
the associated error on the mean for the sample considered in this
work, divided by morphology. $\langle \rm N \rangle$ refers to the number of star 
particles used in the shape computation for the old (ages greater than 12 Gyr) and young (ages less than 2 Gyr) stellar populations.}\label{t:shapes}
\end{deluxetable*}

\subsection{The Intrinsic Shapes of the FIRE\rrrthree{-2} Dwarf Galaxies}\label{s:results:intrinsicshapes}
We present the main results of this section in \autoref{f:baca}. 
In each panel, we show the intrinsic axis ratios of 
the simulated galaxies, B/A and C/A, where $\rm C\leq B\leq A$.
At left,
for reference, we show the well-defined stellar disk and halo 
systems of the high-mass ($M_\odot>10^{10}M_\star$) reference sample
(m11f, m11g, and m12i, see \autoref{t:basic}). 
The intrinsic shapes of young star particles (youngest 25\thh{} percentile) and old star particles (oldest 25\thh{} percentile) are shown in blue and red, respectively. In this panel, the young stars 
clearly are assembled in a well-formed disk ($\rm C<B\sim A$), and the 
old stars occupy a round stellar halo ($\rm A\sim B\sim C$). 
Similarly, in
the middle panel, we show the dwarf galaxies in our sample that have a
clear disk-halo system. These galaxies are characterized by substantial
increases in $\rm C/A$ as a function of age along with relatively little evolution
in $\rm B/A$ compared to $\rm C/A$ over the same comparison, 
and were categorized by visual inspection \fcr{of the galaxies' stellar components}. 
\fcr{Although the oldest (ages greater than 12 Gyr) 
stars may not dominate the light in the outskirts of the FIRE-2 dwarfs, 
we see the same trend towards a more spheroidal shape when comparing the young 
(ages less than 2 Gyr) and intermediate (ages between 2 and 12 Gyr) star 
particles.}

We note that the young stellar disks in these dwarfs
tend to be thicker (higher $C/A$) than the disks in the higher mass galaxies --
this phenomenon is in good agreement with previous observational findings
\citep{padilla2008, sanchezjanssen2010, kadofong2020}. Both the high-mass
and dwarf disks are largely axisymmetric, maintaining a $\rm B/A\sim 1$ in both
their young and old stellar populations. 
Though 
the old stellar population is significantly rounder than the young stellar disks
of these dwarfs, the dwarf stellar halos are
not as round (meaning C/A approaching unity) 
as those of the old stellar population in the high-mass
comparison sample. 
This is again
consistent with the observations of \cite{kadofong2020},
who find that dwarfs
have an average minor-to-major axis ratio $\langle C/A \rangle \sim 0.5$
at 4$\rm R_{eff}$,
significantly flatter than the outer stellar halo of the Milky Way 
\citep[$C/A\gtrsim0.8$,][]{das2016,iorio2018}.
This may be due to a difference in the formation 
mechanism of dwarf stellar halos and massive galaxy stellar halos, which
we will explore in later sections. 

\begin{figure*}[htb]
\centering     
\includegraphics[width=\linewidth]{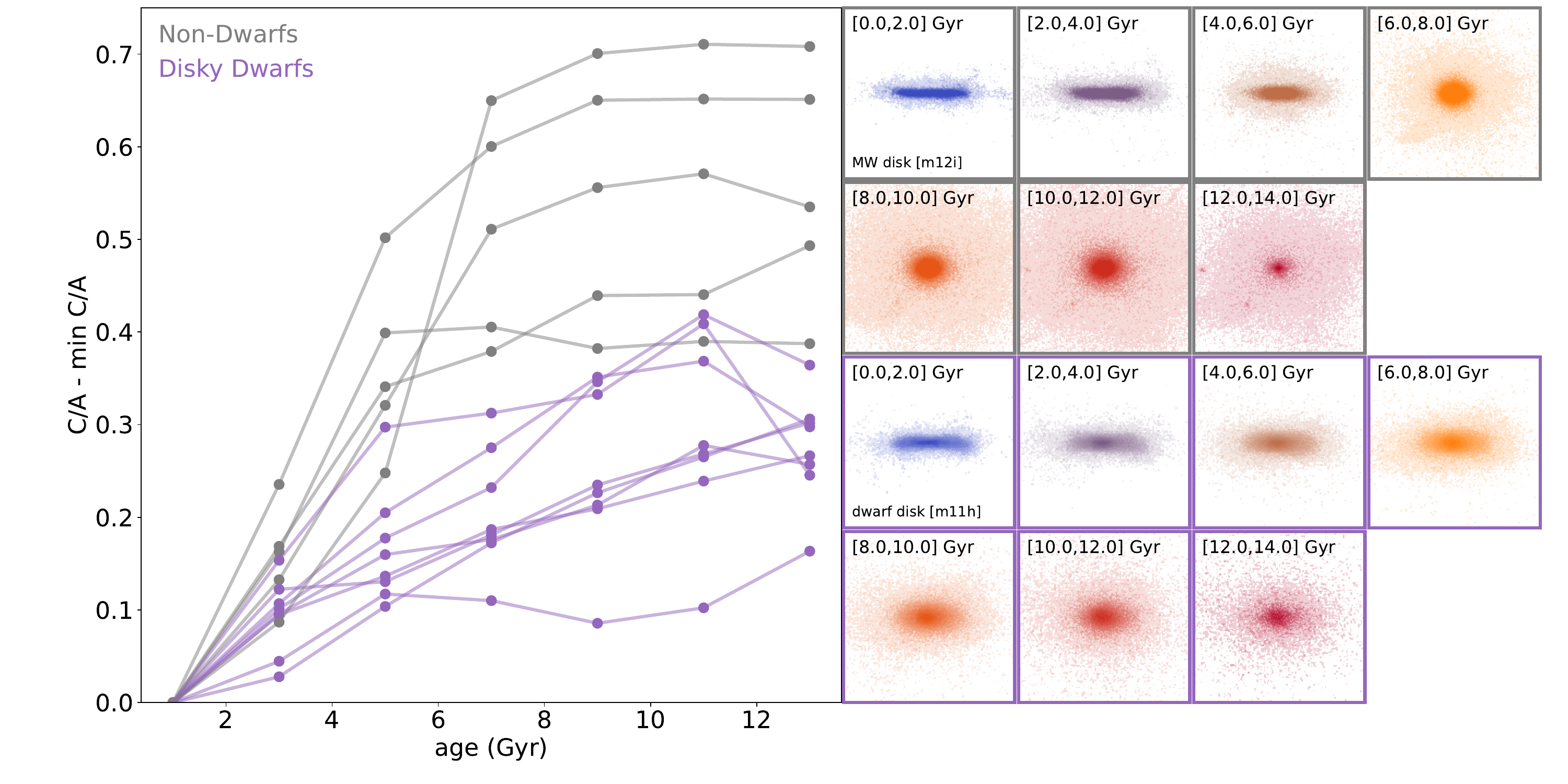}
\caption{
    \textit{At left:} the change in minor-to-major intrinsic axis ratio C/A as a function
    of stellar age, in bins of 2 Gyr. \fcr{We show the change in C/A over the minimum
    C/A across all age bins (because of the monotonic increase in C/A as a function
    of age for these galaxies, this
    is equivalent to the change in C/A over the youngest age bin).} Dwarf disks are shown in purple, while high-mass disks are
    shown in grey. Not only are the old stellar populations of the
    dwarf disks less spheroidal than their high-mass counterparts, d(C/A)/d(age) is markedly
    smaller in the dwarfs than the high-mass galaxies. \textit{At right:} the 
    stellar surface density of an example
    dwarf (top two rows) and massive galaxy (bottom two rows) in bins of 2 Gyr. As at left, while
    the dwarf stellar population thickens gradually with
    increasing age, there is a sharp transition between the
    thin disk and round stellar halo in the more massive galaxy -- this contrast is consistent with the differing origin of
    the dwarf and massive stellar outskirts, wherein dwarf stellar halos are produced via in-situ
    migration of star particles rather than the accretion of satellites. 
    }
\label{f:agebins}
\end{figure*}

The evolution of $C/A$ as a function of age is also
markedly different between the disky dwarfs and high-mass reference
sample. In \autoref{f:agebins}, we show the change in $\rm C/A$ from the
minimum $\rm C/A$ for each disky dwarf (purple) and high-mass galaxy (grey).
Because $\rm C/A$ increases monotonically as a function of age for both these 
groups, this is equivalent to showing the change in $\rm C/A$ from the
$\rm C/A$ of the youngest ($\rm t<2$ Gyr) star particles.
The $x-z$ projections as a function of age for one such disky 
dwarf (top) and one high-mass galaxy (bottom) are shown at right.
The figure serves to illustrate two main points: first, 
the overall change in $C/A$ is systematically smaller for the
disky dwarf sample. Second, the high-mass galaxies approach their
maximum $C/A$ values more rapidly as a function of age, with a sharp
increase in $C/A$ at $5\lesssim t_{\rm age} \lesssim 8$ Gyr and
relatively little change beyond that. This can also be seen qualitatively
in the stark change in shape between the [4,6] Gyr and [6,8] Gyr 
surface densities of the high-mass galaxy. The dwarf galaxies,
meanwhile, increase in $C/A$ gradually as a function of age.

\begin{figure*}[htb]
\centering     
\includegraphics[width=\linewidth]{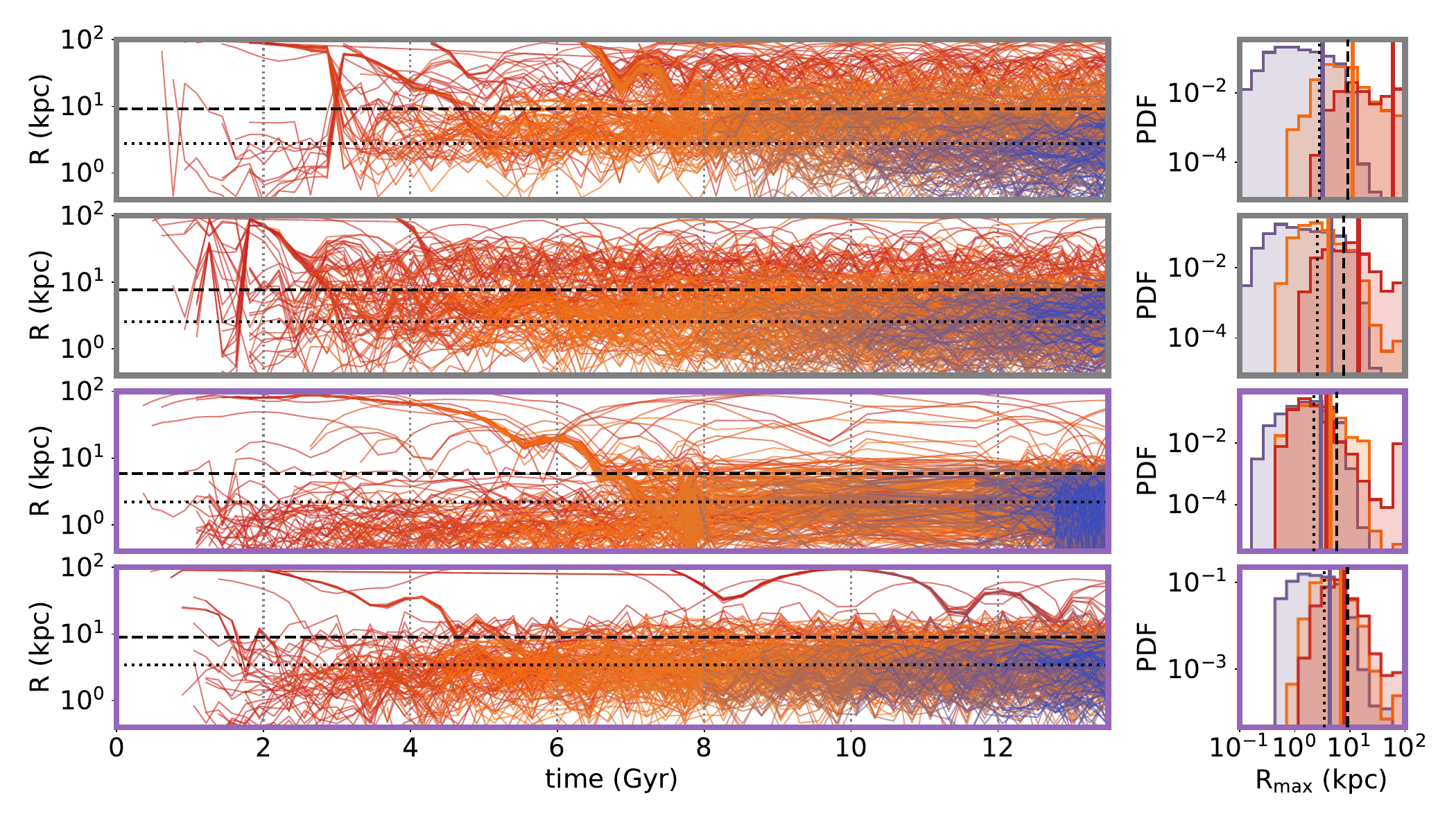}
\caption{\rrrthree{
    \textit{Left:} particle distance from host galaxy center versus 
    time for two high-mass galaxies (m11f (MHD$+$) and m11g (MHD$+$),
    grey-outlined panels) 
    and two disky dwarfs (m11h (MHD$+$) and m11b , purple-outlined
    panels), colored
    by particle age at $z=0$. 
    The dashed (dotted) black lines show the 3D radius that
    contains $90\%$ ($50\%$) the stellar mass at $z=0$.
    \textit{Right:} distribution over maximum distance from
    host achieved for
    star particles younger than 2 Gyr (blue),
    between 6 and 8 Gyr (orange), and older than 12 Gyr (red). 
    Median values for $R_{\rm max}$ for each age component are shown 
    by the colored vertical lines.
    The oldest stars in the massive galaxies are dominated by 
    accreted star particles that remain on large-apocenter orbits to $z=0$.}    
    }
\label{f:ageaccretion}
\end{figure*}

\rrrthree{This difference in the 
three-dimensional shape as a function of
age between the massive and dwarf disk galaxies
can be understand when one considers the origins of the star particles
that constitute these age-binned groups. In \autoref{f:ageaccretion}, 
we show the histories of a random subset of star particles in two massive
disks (top two rows, grey panels) 
and two disky dwarfs (bottom two rows, purple panels).
In the left panels, we show particle distance from
the galaxy center as a function of time, while on the right we show 
the distribution of the maximum distance attained by these particles in 
three age bins (ages $<2$ Gyr in blue, between 6 and 8 Gyr in orange, and
$>12$ Gyr in red). The particle tracks on the
left are colored by age, as are the histograms at right. We find in
particular that
the old, spheroidal components of the massive galaxies (the red tracks and
histograms show particles with ages $>12$ Gyr) correspond to accreted 
components on orbits characterized by large apocenters that persist down to
$z=0$. The young (ages $<2$ Gyr) components are formed in-situ close to the
host center, and the intermediate (ages between 6-8 Gyr) components 
represent a transition between the ex-situ and in-situ-dominated stellar 
populations.}

\begin{figure*}[htb]
\centering     
\includegraphics[width=\linewidth]{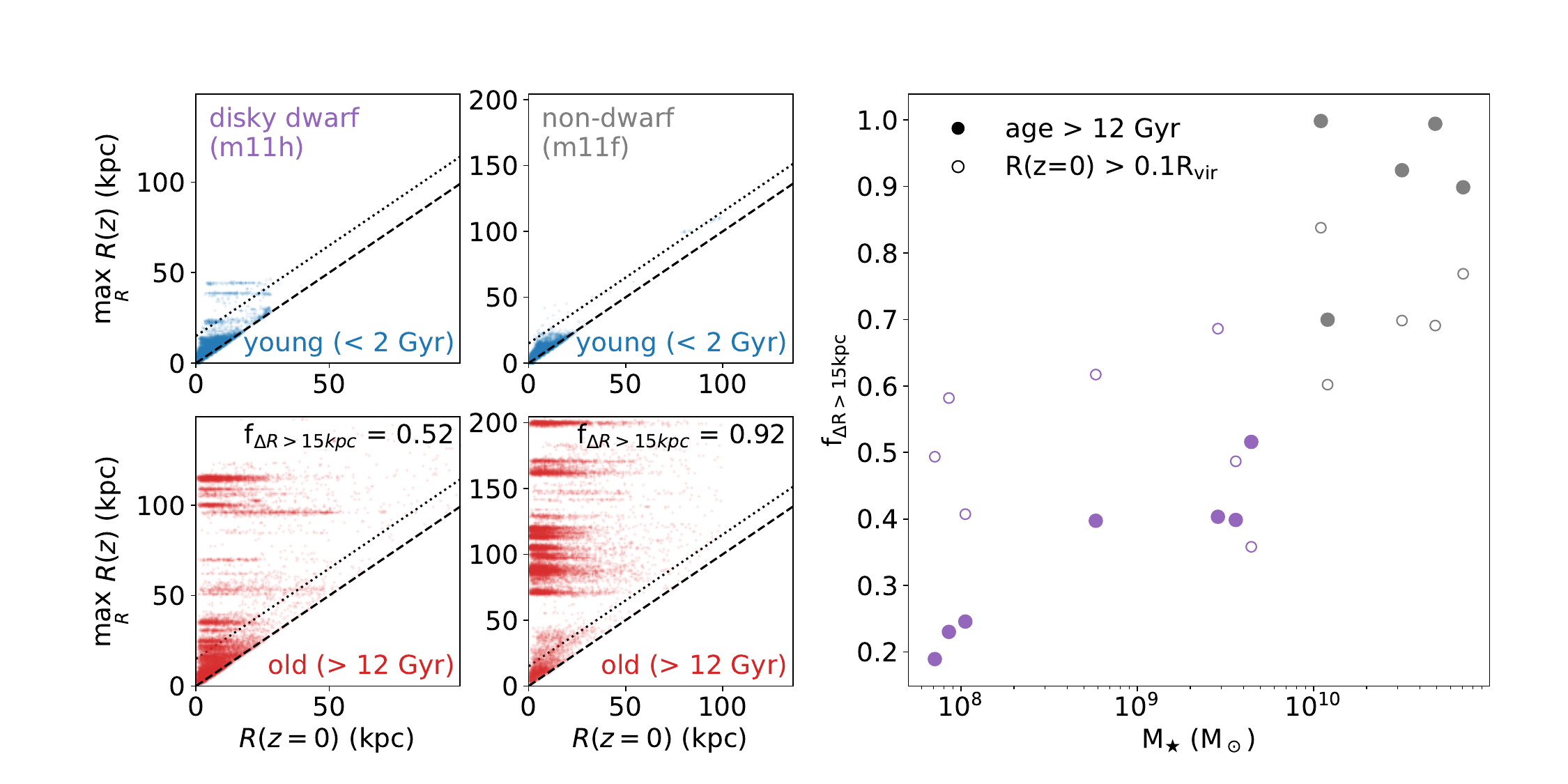}
\caption{ 
    \textit{Lefthand panels:}
    To diagnose the origin of the star particles that populate the stellar halo for the
    galaxies in this work, we plot the maximum distance between the youngest
    (ages $< 2$ Gyr) and oldest (ages $>12$ Gyr) star particles and the host center against their $z=0$ distance. These populations are good tracers of
    the disk and halo structures, if such structures exist. 
    Star particles with an in-situ origin should flare from the 1:1 line (the dashed black line and
    physical minimum), while accreted particles populate horizontal bands
    due to a shared maximum distance that corresponds with
    the original distance between the main host and progenitor system. As a visual aid,
    we show a maximum offset of 15 kpc as the dotted grey line.
    In this space, it is clear that
    while the massive galaxy (m11f, MHD+), 
    is dominated by an accreted stellar halo (bottom right panel), the
    dwarf galaxy (m11h, MHD+), old stellar populations are largely in-situ structures. 
    The x-axis and y-axis limits are set at $\rm 2/3 R_{vir}$ and $\rm R_{vir}$, respectively.
    \fcr{\textit{Right panel:} For the full sample of disky dwarfs (\rrr{m11v (MHD+) is excluded due to a central halo tracking issue at high redshift resulting from an ongoing interaction at $z=0$})  and massive galaxies
    ($\rm M_\star > 10^{10} M_\odot$), the fraction of star particles with a lifetime 
    radial displacement of greater than 15 \rrr{comoving} kpc ($\rm \Delta R = \max_R R(z) - R(z=0)$),
    $\rm f_{\Delta R > 15 kpc}$. We show this quantity both for the old
    star particles (filled circles, ages greater than 12 Gyr) 
    and the star particles with large $z=0$ distances from
    host halo center (unfilled circles, $\rm R > 15$ kpc). Both the old stars and stars in
    the outskirts show an increasing fraction of high-displacement (likely ex-situ) stars
    as a function of stellar mass, in agreement with the picture of an increasingly
    in-situ-dominated stellar halo for dwarf galaxies.}
    }
\label{f:migration}
\end{figure*}

\rrrthree{
The disky dwarfs, meanwhile, are dominated
by stars formed in-situ (though some accreted particles are present) at all 
ages
-- in
the right panels, we show that although the majority of old stars in the massive disks
attain a maximum distance greater than the radius containing 90\% of the 
star particles at $z=0$ ($\rm R_{\star, 90}(z=0)$), 
the star particles in dwarf disks are characterized
by maximum distances less than $\rm R_{\star, 90}(z=0)$ at all ages. That is,
the difference in the evolution of three-dimensional shapes as a function of
age between the massive and dwarf disks is driven by a difference in the
origin of the stars that populate each age group. The spheroidal old component
of the massive disks are driven by an accreted population that fall in on
relatively circular orbits compared to the flatter, in-situ population 
that dominates the old stellar component of the disky dwarfs. We will 
further examine the origin and shape evolution of the disky dwarfs in 
\autoref{s:results:origin} by examining the role of accreted 
star particles in building the stellar outskirts of the full sample 
(\autoref{f:migration}).
}

Finally, in the right panel of \autoref{f:baca}, we show the intrinsic shapes
of the \ndd{} dwarfs in our sample.
\fcr{Although we refer to
these galaxies as ``\ndd{}'' dwarfs, this morphological class refers to galaxies that lack a young stellar disk and do not show a monotonic increase in C/A as a function of
stellar age.}
These dwarfs do not have a well-defined
disk-halo structure, and appear to be significantly different in 
intrinsic shape to \fcr{the population of observed galaxies 
from \cite{kadofong2020}}\footnote{We quantify the possibility of an existent but
minority \ndd{} dwarf population in the observations of \cite{kadofong2020} in 
\autoref{s:toymodel} and discuss this idea further in \autoref{s:discussion:triaxial}}. 
The lack of a well-defined disk in 
particular may point to an overly vigorous or bursty star formation history,
preventing the formation of a rotationally supported disk at low redshift.
\fcr{\cite{elbadry2018quataert} found in particular that the ability of
FIRE\rrrthree{-2} galaxies to form a gaseous disk and maintain a fairly quiescent (less bursty) star
formation history is linked to the accretion of high angular momentum gas at low redshifts,
wherein low-mass FIRE\rrrthree{-2} galaxies struggle to build up stores of this high angular momentum
gas due to both efficient gas removal via star formation feedback and inefficient cooling of
high angular gas in the circumgalactic medium}.

\rrr{While} the star formation feedback and/or cooling prescription
\rrr{may lead to a lack of a disk-halo structure in some dwarfs,} 
\rrr{those} same prescriptions are capable of creating dwarf
stellar disk-halo systems that are remarkably similar in structure to the
observed population. Thus, in this work we will examine the two facets of the
sample separately, examining first the dwarf stellar halo 
(in the simulated galaxies that do form disks) 
and then the origin of the \ndd{} dwarfs.

\fcr{As a final note, we find that 
there is no
change in the morphological classification of the galaxy as a function of the 
physics implementation for the majority of the dwarfs in the sample. 
That is, it appears that the $z=0$ stellar intrinsic structure is largely 
unaffected by the additional physics implemented in the (MHD+) and (CR+) runs 
as described in \autoref{s:methods:fire}. However, we do find that two galaxies, m11i
and m11v, are disky in their (MHD+) runs and non-disky in their (CR+) or (Hydro+) runs.
\rrrthree{We find that the change in morphological classification
in m11i is due to its proximity to the boundary between 
our disky and non-disky classifications -- indeed, in the middle panel
of \autoref{f:baca}, the disky m11i (MHD+) galaxy is shown by the pair
of points at the lowest $B/A$ value of the dwarf sample. Thus, we 
conclude that the change in morphological class of m11i is due to the
rigid nature of the binary classification that we use in this work. 
Though the disky and non-disky dwarfs are generally well-separated 
morphological classes, there do exist edge cases such as m11i in which
the classification can change as a result of relatively small 
perturbations in intrinsic axes space.
m11v does undergo a significant shape change between the
MHD+ ($B/A=0.90$ and $C/A=0.46$ for $t<2$ Gyr stars) and 
CR+ ($B/A=0.25$ and $C/A=0.13$) runs. However, m11v is also 
experiencing an ongoing interaction that is the likely cause of this
shape variation.}}

\begin{figure}[htb]
\centering     
\includegraphics[width=\linewidth]{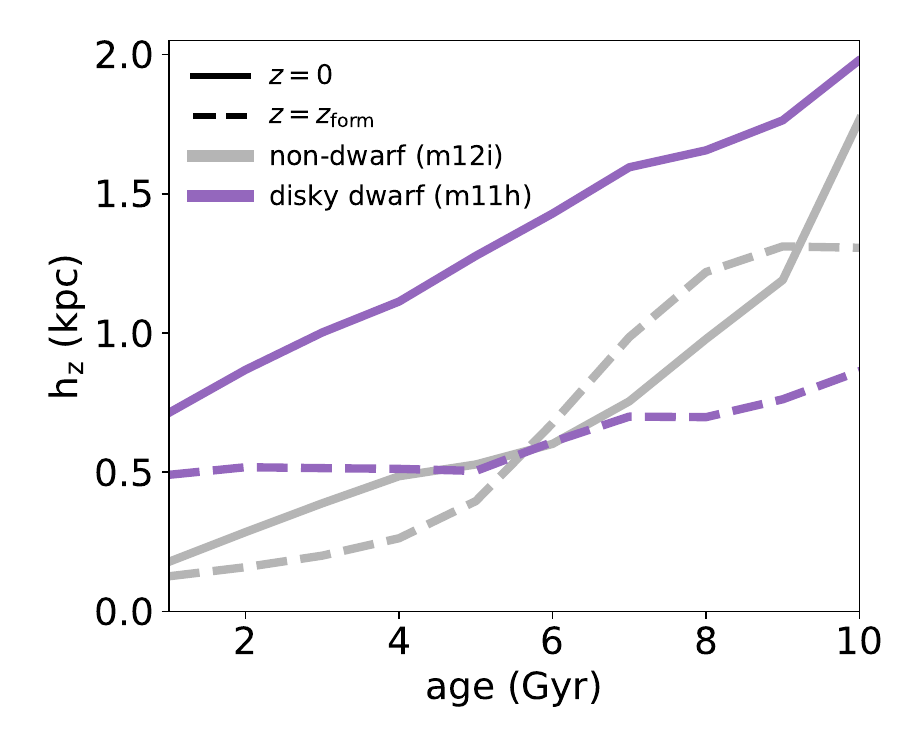}
\caption{ 
    \fcr{For a massive (grey, $\rm M_{halo} \sim 10^{12} M_\odot$) and 
    dwarf (purple, $\rm M_{halo} \sim 10^{11} M_\odot$) disk galaxy, we show
    the change in the exponential stellar scale height $\rm h_z$ as a function
    of age. \rrr{For the star particles with a given age, t}he solid curves show the current, $\rm z=0$ scale height \rrr{of those star particles}, while the
    dashed curves show the scale height \rrr{of the same single-aged population}
    at $z=z_{\rm form}$\rrrthree{$=z(t_{lb}=\rm age)$}. 
    We exclude stars older than 10 Gyr, as in the massive galaxy
    they are likely to be dominated by ex-situ stars
    which should not be well-described by an 
    exponential distribution at birth.
    Whereas the 
    stars in the massive galaxy show a negligible change in vertical scale, the dwarf stars are consistently
    more extended at current times than they were at their formation time.
    This
    is consistent with the picture in which the dwarf stellar outskirts are 
    populated by stars that have been heated to larger radii after
    their formation. In contrast, the
    massive galaxy FIRE-2 thick disks are formed due to an overall change in the configuration
    of star formation as a function of cosmic time (see \citealt{yu2021}).}
    }
\label{f:hzevolution}
\end{figure}

\begin{figure*}[htb]
\centering     
\includegraphics[width=\linewidth]{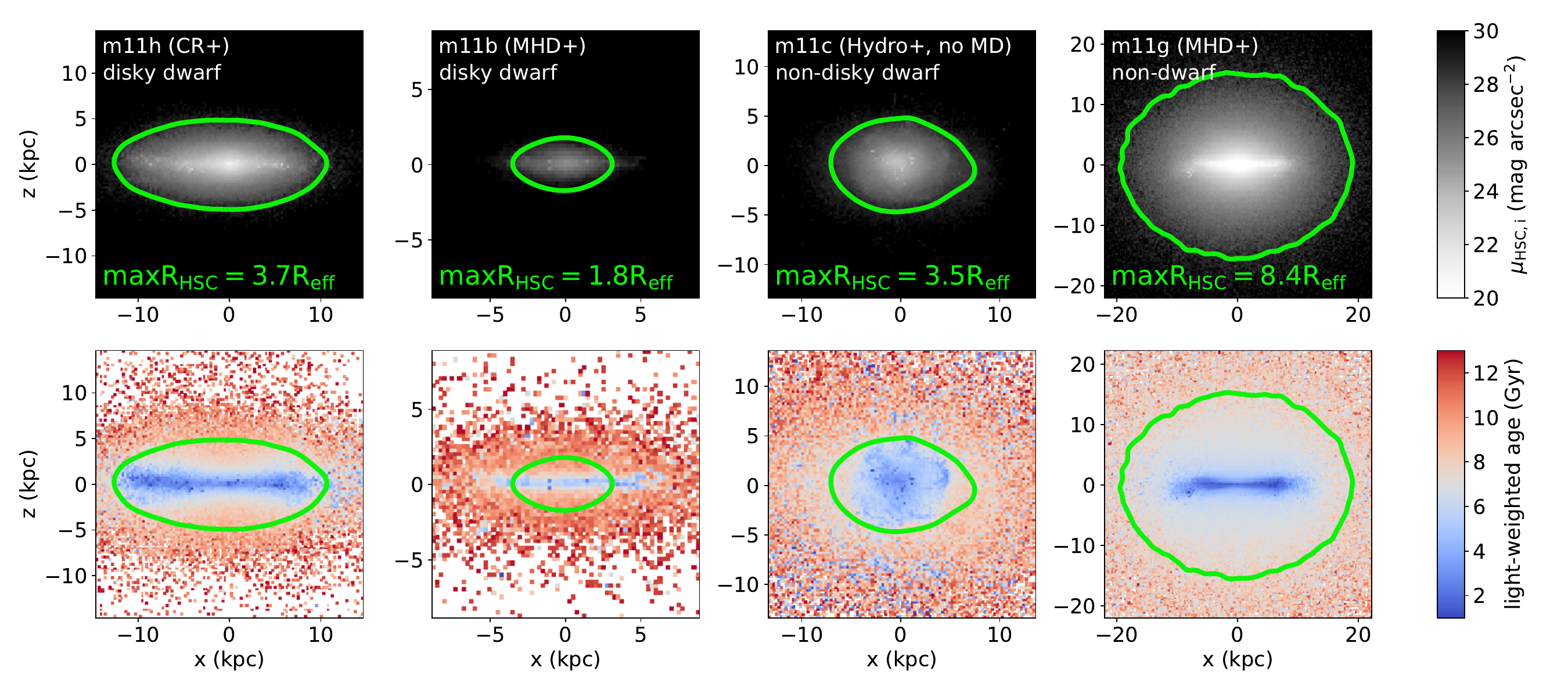}
\caption{ 
    \rrr{Mock HSC-i band surface brightness maps (top row) and HSC-i light-weighted age maps (bottom row) for an example set of galaxies which span the mass and
    morphologies studied in this work. From left to right, we show: m11h (CR+), a 
    disky dwarf at the high end of the dwarf stellar mass range ($\rm M_{\star,90}=3\times10^9M_\odot$), m11b (MHD+), a disky dwarf at the 
    low end of the dwarf stellar mass range ($\rm M_{\star,90}=7\times10^7M_\odot$), m11c (Hydro+, no MD), a non-disky dwarf,
    and m11g (MHD+), a massive galaxy ($M_{\star,90}=5\times10^{10}M_\odot$). The
    lime contour in each panel is an isophote at $\rm \mu_i=28.5$ mag arcsec$^{-2}$, the
    approximate surface brightness limit of the HSC-SSP i-band for galaxy outskirts.
    The values written in lime at the bottom of the top row of panels show the
    approximate maximum radius that HSC-SSP observations would be able to reach
    for each galaxy, as a function of the i-band effective radius of
    the galaxy (as determined
    by a single two-dimensional \sersic{} fit). }
    }
\label{f:observability}
\end{figure*}

\begin{figure}[htb]
\centering     
\includegraphics[width=\linewidth]{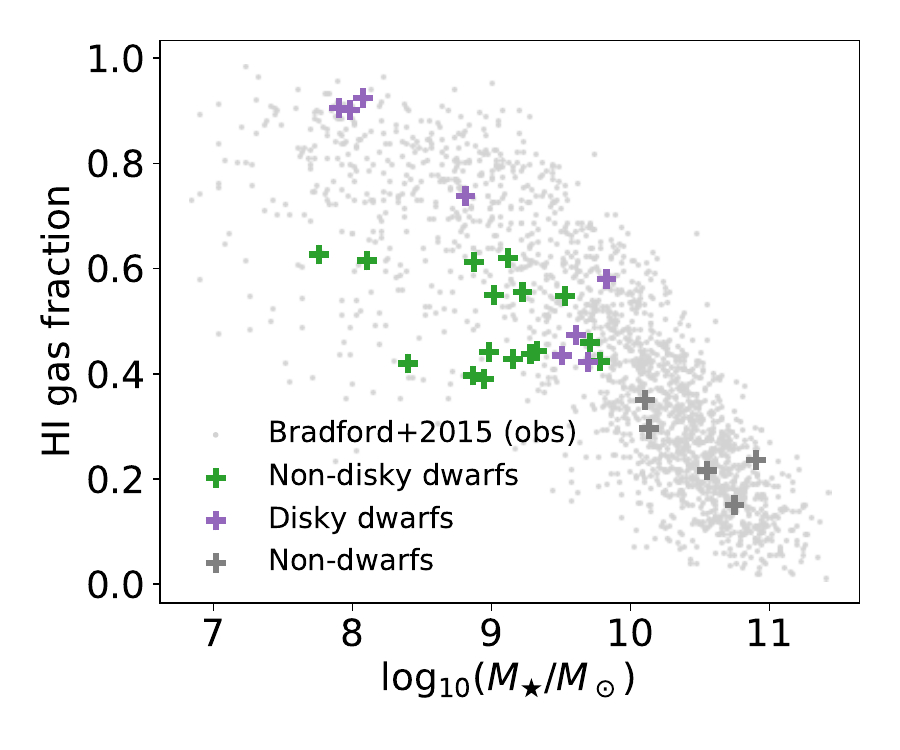}
\caption{ 
    The HI gas fraction $M_{\rm HI}/(M_\bigstar+M_{\rm HI})$ versus stellar mass for the
    simulated FIRE\rrrthree{-2} galaxies (marked as crosses) 
    and the observed sample of \cite[][light grey scatter]{bradford2015}.  The FIRE\rrrthree{-2} galaxies are
    colored by morphology, as indicated. The FIRE-2 dwarfs tend to
    be overly HI-depleted compared to observed galaxies \citep[see also][]{elbadry2018}, 
    so the gas rich end of the FIRE\rrrthree{-2} dwarfs are
    the most consistent with observations. In particular, the galaxies that successfully 
    form disks have HI gas fractions consistent with the observed relation, while the \ndd{}
    dwarfs are systematically gas depleted relative to the \cite{bradford2015} sample.
    }
\label{f:bradford}
\end{figure}

\begin{figure*}[htb]
\centering     
\includegraphics[width=\linewidth]{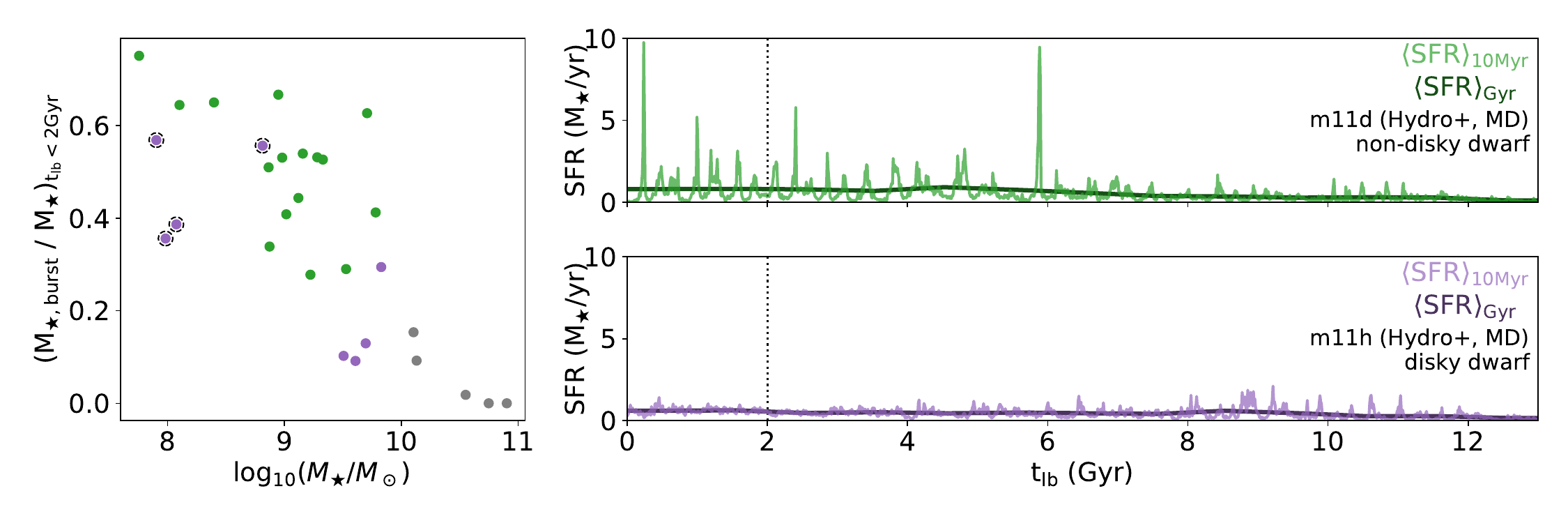}
\caption{ 
    At left, the fraction of stellar mass built up in bursts (defined as time periods where
    $\rm \langle SFR\rangle_{10Myr}/\langle SFR\rangle_{Gyr} > 1.5$) versus 
    stellar mass for the sample of FIRE\rrrthree{-2} dwarfs. The purple points
    show disky dwarfs, while the green points show the \ndd{} dwarfs, \rrr{and the grey points show the
    non-dwarfs}. The low-mass dwarfs
    that have HI gas fractions of greater than $(M_{\rm HI}/(M_\bigstar+M_{\rm HI})=0.7$ are
    shown with a dashed outline. 
    We find that the dwarfs
    that are able to form a young stellar disk have either particularly quiescent (non-bursty) recent
    star formation histories, or are particularly gas rich relative to the other
    simulated galaxies. At right we show the star formation history for a 
    \ndd{} dwarf (top, m11d \rrrthree{(Hydro+, MD)}) and a disky dwarf (bottom, m11h \rrrthree{(Hydro+, MD)}). 
    In each panel we show the 10 Myr-averaged and
    Gyr-averaged star formation rates as the light and dark curves, respectively. We also show
    vertical lines at $\rm t_{lb} =2$ Gyr, the time period over which the burst mass fraction is
    computed. 
    }
\label{f:burstgas}
\end{figure*}

\subsection{The Origin of Dwarf Halo Stars}\label{s:results:origin}
In \autoref{f:migration}, we show for a selection of galaxies
the maximum distance of the star particles from the host halo center since $z\sim 6.2$ as a function of the present-day distance. 
This figure acts to visualize the contributions to
the young (top, blue) and old (bottom, red) stellar populations by in-situ
and ex-situ components:
in-situ star particles flare out from the 1:1 line, 
while accreted star particles occupy horizontal bands that correspond to the
maximum distance of the progenitor system over the time considered.
\rrrthree{In the set of panels at left we show the $z=0$ distances versus 
maximum distance attained from host center for a disky dwarf (m11h (MHD+)) and
non-dwarf (m11f (MHD+)), while at right we show the fraction of particles that
attain a displacement of greater than 15 kpc (i.e. those particles
that lie above the dotted line in the left panels) for the full set of 
disky dwarfs and massive galaxies. The right panel shows this metric for both
the old stars (filled points) and for all particles with a $z=0$ host distance of
greater than 10\%$R_{\rm vir}$.}

As expected, the young stellar populations in all the galaxies shown 
in \autoref{f:migration} are dominated by in-situ contributions.
However, the difference
in how the dwarfs and the MW-mass galaxy have assembled their old stellar
populations is apparent. Whereas the old star particles 
in the dwarfs (and therefore their stellar halos) are dominated by 
in-situ star particles, the old stellar population in the MW-mass galaxy
is completely dominated by accreted star particles. 
We find no significant difference between the accreted halo fraction of the \ndd{} 
dwarfs and disky dwarfs, though the sample size is too small to straightforwardly 
extrapolate this behavior to the general dwarf population.

\rrr{The diagnosis of
halo star particle origin that we use in this work is not meant to produce exact ex-situ fractions, but is
instead aimed to identify bulk changes in halo formation mechanism. We test our
$\Delta R > 15$ \rrrtwo{comoving} kpc threshold against a cut in formation distance, which has been
used in prior work with the FIRE\rrrthree{-2} simulations to track ex-situ contributions
\citep{sanderson2018} and has been showed to produce accreted fractions in 
agreement with subhalo tracking from merger trees for MW-like galaxies in 
FIRE-2 \citep{necib2019}. We compare these two ex-situ flagging methods for the disky dwarf m11h (Hydro+, MD) and
the high mass galaxy m12i (Hydro+, no MD),
and find that our $\rm \Delta R$ cut produces accreted fractions consistent to
within 10\% of those based on a cut on formation distance. We
also consider the effect that our choice of a 
threshold cut at $\Delta \rm R=15$ kpc plays on our estimated
ex-situ fractions, and find that a cut of $\Delta \rm R=30$ kpc \rrrtwo{($\Delta \rm R=7$ kpc)} changes the estimated ex-situ fractions by no more
than 12\% \rrrtwo{($5.7\%$)}, and does not affect the trend with stellar mass shown in \autoref{f:migration}.}
We also find that this trend holds when we consider all stars in the outskirts, regardless of age ($\rm R(z=0)>0.1 R_{vir}$, \rrr{open points in \autoref{f:migration}}). 
Indeed, the divergent behavior seen in \autoref{f:migration} between the dwarf and massive galaxy stellar
halos is 
consistent with the supposition that the stellar halos of massive galaxies
are dominated by ex-situ stars, while the
dwarf stellar halos are largely in-situ structures.

\rrr{The thick and in-situ nature of the dwarf stellar outskirts in 
FIRE-2 also draws a clear structural similarity to the thick disks in more
massive galaxies. Though the origin, existence, and nature of thick disks in
the Milky Way and beyond is still a matter of significant debate
\citep[see, e.g.][]{yoachim2006, bovy2012, belokurov2020, agertz2021,park2021},
it is straightforward to make an internal comparison between the dwarf 
stellar outskirts and massive thick disks within the context of the FIRE-2 
simulations.
It has been demonstrated that the thickened shape of the 
thick disks in the more massive FIRE-2 galaxies is \rrrthree{caused by
star formation characterized by larger vertical scale heights}
during bursty phase\rrrthree{s} of star formation
at lookback times of $t_{\rm lb} \gtrsim 5$ Gyr \citep{yu2021}. 
It is thus of interest to examine whether the shape of the FIRE-2 dwarf
stellar populations are set by stellar migration or by a change in the
star formation configuration as a function of cosmic time. To address this,
we show in \autoref{f:hzevolution} the \rrrthree{difference in the scale 
height of the star particles in bins of age
at $z=0$ and at $z=z_{\rm form}$, where $z_{\rm form}$ is the formation redshift
(i.e. the lookback time equivalent to the $z=0$ age of the star particle population),
of star } 
particles in a disky dwarf galaxy (m11h) and a massive galaxy (m12i) 
as a function of star particle age.}
\rrrthree{We compute scale heights from an exponential fit to the vertical stellar
surface density in age bins spaced by 1 Gyr with widths of 2 Gyr, and 
exclude
star particles with ages greater than 10 Gyr for the massive galaxy in 
this figure, 
as the massive galaxies are expected to be dominated by ex-situ stars at this age (see 
\autoref{f:migration}) and the majority of the massive FIRE-2 thick disks
are assembled at times more recent than 10 Gyr \citep{yu2021}. }


\rrr{We first consider the massive 
FIRE-2 galaxy; \autoref{f:hzevolution} demonstrates that the 
star particles in this galaxy show a negligible change in
vertical scale height between $\rm z=z_{form}$ and $\rm z=0$. 
This lack of scale height evolution is 
consistent with the results of \cite{yu2021}, who find that the thick disks
in these massive galaxies are formed as thick disks from birth, rather than
from stellar displacement after formation. If, however, we turn our 
attention to the dwarf disk, we see that the $z=0$ vertical scale height
is consistently higher than the vertical scale height at formation. This
evolution is instead consistent with the picture of \cite{elbadry2016}, 
where the stellar outskirts are indeed formed from the displacement of
stars after formation. This implies that the dwarf stellar outskirts
in FIRE-2 are distinct in formation mechanism 
from both the mechanisms that create 
massive stellar halos and massive stellar thick disks in the 
FIRE-2 simulations.}

\rrr{\subsection{The Observability of the FIRE-2 Dwarf Stellar Halos}}\label{s:results:observability}

\rrr{As discussed in \autoref{s:methods:intrinsicshape}, the way in which 
we measure the intrinsic shape of the FIRE-2 dwarfs is significantly different
than the measurement methods applied to observations in \cite{kadofong2020}. Thus,
it is of interest to quantify the extent to which the stellar halos around the
FIRE\rrrthree{-2} dwarfs would be observable in Hyper Suprime-Cam Subaru Strategic Program (HSC-SSP), the imaging survey used by \cite{kadofong2020}.
This test will help to understand the extent to which the FIRE-2 dwarf stellar halos and
observed dwarf stellar outskirts track the same physical structure. }

\rrr{
In order to quantify the observability of the FIRE-2 dwarf stellar halos, we
generate mock observations of the sample in this work using the Flexible Population 
Synthesis (FSPS) package \citep{conroy2009,conroy2010} to generate simple
stellar populations over a grid of stellar ages (between $10^5$ and $10^{10.3}$ yr) and metallicities (in the range $-4<\log_{10}(Z/Z_\odot)<1$). We then compute the 
stellar mass-to-light ratio in the HSC i-band for each model spectrum. Each star particle
in the simulated dwarfs is then assigned an HSC i-band mass-to-light ratio based on its age and metallicity via 
linear interpolation of the model grid; this process allows us to compute 
surface brightness and light-weighted age maps for each galaxy in our sample.
Following \cite{sanderson2018}, we construct these maps via a simple binning
of star particles with a box size of 0.25 kpc on a side.
}

\rrr{
We show these surface brightness and light-weighted age maps in \autoref{f:observability} for a set of galaxies that span the stellar masses
and morphologies of the sample at hand. Each panel shows the isophote
at the nominal HSC i-band
surface brightness limit for galaxy profiles ($\rm \mu_i=28.5$ mag arcsec$^{-2}$)
as a lime green contour. In order to best quantify the ability of HSC to detect 
the old, round stellar outskirts of the FIRE-2 galaxies, we project each
galaxy along the y-axis. We also compute an effective radius in the HSC i-band 
using this projection via a two-dimensional, single-component \sersic{} fit; in
each panel, we note the maximum radial distance of the $\rm \mu_i=28.5$ mag arcsec$^{-2}$ isophote in terms of the HSC i-band effective radius, which
we will denote as $\rm \max R_{HSC, obs}$.
}

\rrr{Naturally, we find that the 
value of $\rm \max R_{HSC, obs}$ increases as a function of stellar mass;
adopting the stellar mass bins used in \cite{kadofong2020}, we find
that the FIRE-2 galaxies with stellar masses of $\rm M_\star < 10^{8.5} M_\odot$ 
have a mean (standard deviation of) 
$\rm \max R_{HSC, obs}$ of $\rm 2.0\ (0.2)\ R_{\rm eff, HSC-i}$, while the
galaxies with stellar masses of $10^{8.5} < \rm M_\star < 10^{9} M_\odot$ 
have a mean (standard deviation of) 
$\rm \max R_{HSC, obs}$ of $\rm 3.6\ (0.8)\ R_{\rm eff, HSC-i}$, and the
galaxies with stellar masses of $10^9 < \rm M_\star < 10^{9.6} M_\odot$ 
have a mean (standard deviation of) 
$\rm \max R_{HSC, obs}$ of $\rm 4.1\ (1.7)\ R_{\rm eff, HSC-i}$. These
values are consistent with the spatial limit of the HSC imaging for the dwarfs in \cite{kadofong2020}.
}

\rrr{
As shown in \autoref{f:observability}, in the cases where a young stellar disk is
formed (i.e. excluding the non-disky dwarfs due to the lack of an analogous 
population in the observations), the mock HSC observations extend beyond the young
stellar disk into the rounder and older stellar outskirts. This finding supports
our initial assertion in \autoref{s:methods:intrinsicshape} that the 
age-separated intrinsic shape measurements in the simulated galaxies and the
radially-separated intrinsic shapes inferred from the observed dwarf population
are tracing the same physical structure.
}

\subsection{Properties of the Disky and \NDD{} Dwarfs}
Though diagnosing, in detail, the formation path of \ndd{} dwarfs
is not the main aim of this paper, it is informative to compare their 
global properties to those of the disky FIRE\rrrthree{-2} dwarfs and to observed 
dwarf galaxies. 

\fcr{It is known that the dwarfs in FIRE\rrrthree{-2} are,
on average, somewhat lower in their HI gas fractions as compared to observed dwarfs
\citep{elbadry2018bradford}. Here we expand upon this trend to demonstrate that
the gas content of the FIRE-2 dwarfs are also linked to their three dimensional
structure.}
To do so, we compare the 
HI gas content\footnote{\rrrtwo{The ionized fraction of hydrogen is calculated self-consistently from the cooling routines implemented in FIRE-2 -- a succinct description of the cooling function and radiative feedback processes included in FIRE-2 can be found in appendices B and E of \cite{hopkins2018_main}, respectively.}} 
of the FIRE\rrrthree{-2} simulations to that of the
observed sample of \cite{bradford2015}. \rrr{We compute the HI and stellar masses
within 200 kpc of the galaxy center to derive the HI gas fractions of the FIRE-2
galaxies (though we note that a choice of 50 kpc from galaxy center would change
the HI gas fractions by a few percent at most, and generally much less).}
Drawn from Sloan Digital Sky Survey (SDSS)
spectroscopically confirmed targets with Arecibo Legacy Fast ALFA (ALFALFA) 21cm
observations, this sample consists of isolated galaxies with stellar masses 
$10^7\lesssim M_\star/M_\odot \lesssim 10^{11}$. We compare this isolated HI sample
to the FIRE\rrrthree{-2} galaxies in \autoref{f:bradford}; the observed galaxies are shown in 
grey scatter, while the simulated galaxies are shown by the crosses (colored by morphology,
as indicated). Notably, while the disky dwarfs lie on the main relation found by
\cite{bradford2015}, the \ndd{} dwarfs are systematically more gas-poor at fixed 
stellar mass than the observed galaxies. 

Next, we contrast the star formation histories (SFH) of the \ndd{} and 
disky dwarfs. In \autoref{f:burstgas}, we show at left the fraction of stellar mass 
built in starbursts over the last 2 Gyr versus the galaxy's stellar mass. We define a starburst as the snapshots for which
\begin{equation}
    \frac{\rm \langle SFR\rangle_{10Myr}}{\rm \langle SFR \rangle_{1Gyr}} > 1.5.
\end{equation}
We find that the FIRE\rrrthree{-2} dwarfs that succeed in forming disks are
those that are either particularly gas rich (as seen in \autoref{f:burstgas}; for
easy comparison we outline the galaxies where the HI gas fraction exceed 0.7 
with a dashed circle) or have particularly non-bursty
recent star formation histories for their stellar mass. 
To visualize the difference in star formation histories, we show the SFH 
for a \ndd{} dwarf at top (green), and a 
disky dwarf with a low  burst fraction at bottom (purple). Though both are highly bursty
at high redshift, the disky dwarf becomes significantly less bursty at low redshift.

\section{Discussion}\label{s:discussion}
In the previous section, we showed that the FIRE\rrrthree{-2} dwarfs host stellar
outskirts that are assembled from in-situ star particles, a distinctly
different formation pathway than the stellar halos of more massive galaxies.
Here, we place our results into the context of simulations of dwarf
galaxies at large, and compare the simulated dwarf properties to those
of observed dwarf galaxies.

\subsection{Comparison to Other Simulations}\label{s:discussion:comparison}
As the FIRE\rrrthree{-2} simulation suite are a set of cosmological zooms, we
will first contextualize our results via comparison to similar
studies done on cosmological and non-cosmological
simulations.

Because of the high resolution necessary to study the stellar outskirts of
low-mass halos, there are relatively few cosmological simulations
that make similar structural measurements on dwarf galaxies to which
we can compare. \citet{pillepich2019} use TNG50, the highest resolution volume of the 
IllustrisTNG simulations \citep{pillepich2018}, to present mass-weighted
intrinsic shape measurements made in an ellipsoidal aperture
at twice the half-mass radius, $\rm 2R_{1/2, \rm mass}$, with a thickness
of $\rm 0.4R_{1/2,\rm mass}$. 
TNG50 reaches a baryonic
mass resolution of \rm $8.5\times10^4 M_\odot$ and a DM mass resolution of
$\rm 4.5\times10^5 M_\odot$ (about an order of magnitude lower resolution in both
baryons and dark matter than the galaxies studied in this work)
\fcr{and a minimum gravitational force softening of 74 pc for
the gas (compared to $\sim2$ pc for the galaxies in this work)}. TNG50 contains approximately $\sim5500$
dwarf galaxies with stellar masses in
the range $10^8 < M_\star/M_\odot < 10^{10}$. The 
intrinsic shape of the dwarf galaxies in
TNG50 show a strong dependence on stellar mass, with the 
$10^9 < M_\star/M_\odot < 10^{10}$ 
galaxies characterized by thick disks and the 
$10^8 < M_\star/M_\odot < 10^{9}$ dwarfs significantly more spheroidal. 
We see no such trend with mass \rrr{in the FIRE\rrrthree{-2} simulation suite}.
It is of interest that the high-mass TNG50 dwarfs form 
thick disks, even in the old stellar population. 
This is in contrast to the FIRE\rrrthree{-2} dwarfs 
that succeed in forming thick disks; the FIRE\rrrthree{-2} stellar disks are
dominated by young stars, while the old stellar populations that
dominate the stellar mass budget maintain
largely spheroidal shapes at all radii (for cumulative measurements). 
Observations indicate that high-mass dwarfs should generically
host stellar disks, as they do in TNG50 \citep{kadofong2020}, but 
also find morphological differences as a function of age in resolved
star studies, as is seen in FIRE\rrrthree{-2} \citep{zaritsky2000,aparicio2000,aparicio2000a,hidalgo2003, demers2006,bernard2007,stinson2009,strader2012,nidever2019a,nidever2019,pucha2019}.

It is also informative to compare to higher resolution, non-cosmological
simulations of dwarf galaxies. In particular, \cite{smith2020} 
examine the effect of
photoionization and photoelectric heating from young stars on the efficacy of supernova
feedback and the resultant impact on the SFH and disk structure. 
They find that the inclusion of photoionization significantly dampens
the burstiness of the simulated dwarf SFH, and that runs that
include only SNe feedback result in superbubbles that significantly
disrupt the gaseous disk. It is important to note that the
galaxy studied in \cite{smith2020} is initialized with both a stellar
and gas disk, and is
significantly lower in stellar
mass ($M_\star\sim10^7M_\odot$) than the galaxies studied in this work;
it is observationally uncertain whether galaxies in this mass range 
are generally disky \citep{kadofong2020, carlsten2021}. 
However, their finding that 
over-vigorous star formation feedback can disrupt the disk is in line with
our finding that it is the bursty and gas poor FIRE\rrrthree{-2} dwarfs that do not
form young stellar disks.

\rrr{Finally, it has been noted that dark matter particles may induce 
numerical heating in the stellar component of simulated galaxies. We 
do not expect this numerical heating to significantly affect our results
-- the resolution of the FIRE-2 dwarfs surpasses the dark matter particle
mass limit estimated by \cite{ludlow2021} at which numerical heating 
accounts for no more than 10\% that of the virial velocity
($\rm m_{DM} \lesssim 4.5\times10^4 M_\odot$; the maximum particle mass
of the FIRE\rrrthree{-2} dwarfs is $\rm m_{DM} = 3.8\times10^4 M_\odot$). Indeed, 
for the massive FIRE\rrrthree{-2} galaxy m12i, \cite{ludlow2021} estimates that the
change in scale height attributable to numerical heating is only 70 parsec, 
which is much smaller than the change in scale height observed in the FIRE
dwarfs (see \autoref{f:hzevolution}). Additionally, a study of the
somewhat smaller ($\rm M_{halo} \sim 10^{10} M_\odot$) FIRE\rrrthree{-2} galaxies run at
30$\rm M_\odot$ resolution 
showed that structural measurements in dwarfs at fiducial FIRE\rrrthree{-2} resolution are robust to
resolution effects \citep{wheeler2019}.}

\subsection{A Divergent Path of Low-mass Stellar Halo Formation}\label{s:discussion:formation}
Due to the steepening stellar-to-halo mass relation at low masses, it has 
long been thought unlikely for dwarf galaxies to assemble a significant stellar halo
population via the accretion of satellite galaxies. Thus, the 
observed presence of extended old \citep{lin1983,minniti1996,minniti1999,zaritsky2000,aparicio2000, aparicio2000a,hidalgo2003, demers2006, bernard2007, stinson2009, strader2012, nidever2019, nidever2019a, pucha2019} and round \citep{kadofong2020}
stellar outskirts around dwarf galaxies points to a different mode of
stellar halo formation at low masses.

In this work, we have shown that the FIRE-2 dwarf galaxies are capable of
producing an extended old and round stellar population in conjunction with a young
stellar disk, and we find that the stellar outskirts of the FIRE\rrrthree{-2} dwarfs 
reproduce several key qualities of observed high-mass
dwarf stellar halos. \rrrthree{However, FIRE-2 also produces a large 
contingent of non-disky dwarfs whose structural and gas properties are 
at odds with the observed population. 
Nevertheless, it is informative to consider the 
mechanism through which
the high resolution and cosmological dwarfs simulated in FIRE-2
form disk-halo systems, 
and to
consider the plausibility of such a mechanism in the context of 
our knowledge of the observed dwarf population.}

First, in a marked contrast to the
ex-situ stellar halos of massive galaxies, these dwarf stellar halo populations are 
dominated by in-situ stars that occupy an increasingly spheroidal component as
a function of age. This behavior explains the apparent
ubiquity of round stellar outskirts around high-mass dwarfs ($M_\star>10^{8.5}M_\odot$),
as feedback should be a generic process in all galaxies that proceeds regardless
of accretion history or environment.
The existence of an old, in-situ population in the
outskirts of the FIRE\rrrthree{-2} dwarfs is not surprising, as it has been shown that
the radial migration of stars due to baryon-driven potential fluctuations 
operates most strongly in FIRE\rrrthree{-2} dwarfs in this mass range \citep{elbadry2016, graus2019}. This
feedback-driven migration has previously been suggested as a formation mechanism
for in-situ stellar halos \citep{stinson2009, maxwell2012}. \rrr{Star formation
feedback-driven size fluctuations have also 
recently been found via correlations between size and 
star formation rate in both the FIRE-2 dwarfs and a sample of nearby dwarfs
from the Local Volume Legacy Survey \citep{emami2021}.}

Furthermore, the old stellar population of the FIRE\rrrthree{-2} dwarfs, though largely spheroidal
compared to the young stellar disk populations, are 
 significantly more flattened (a lower minor-to-major intrinsic axis ratio
$C/A$) than are the halos of the higher mass FIRE\rrrthree{-2} galaxies. This is again
due to their in-situ formation mechanism, wherein the disk gradually thickens
into the halo as the stars age. 
These preferentially flatter dwarf
stellar halos are in good agreement with observations
of dwarf stellar outskirts, which find that dwarf stellar outskirts at 4 effective
radii are preferentially flatter than the Milky Way at the same distance.
Indeed, the mean $\rm C/A$ of the young stellar population of disky dwarfs in FIRE
($\langle \rm C/A \rangle =0.21\pm0.03$, where $\langle \rangle$ denotes the sample mean,
see \autoref{t:shapes}) 
is similar to the maximum a posteriori estimate of the mean $\rm C/A$ 
of the massive dwarf population inferred
by \cite{kadofong2020} ($\rm\langle C/A\rangle = 0.30 \pm 0.01$ for dwarfs of stellar mass 
$\rm 10^{9}<M_\star/M_\odot <10^{9.6}$ and $\rm\langle C/A\rangle = 0.32 \pm 0.02$ 
for dwarfs of stellar mass 
$\rm 10^{8.5}<M_\star/M_\odot <10^{9}$, see Table 1 of \citealt{kadofong2020}). 
However, we stress that because of the differences in the intrinsic
shape measurement method \fcr{(differential versus cumulative) and
the exact stellar populations being measured (i-band light-weighted versus
age-separated)}, we are not emphasizing here the apparent 
agreement in the numerical value of $\rm \langle C/A\rangle$ between the
observed and simulated dwarfs. Rather, we point to the difference in
$\langle C/A \rangle$ of the simulated (observed) dwarfs and 
simulated (observed) high-mass galaxies as the key observational
constraint on the difference in 
the assembly mode of their stellar halos.

\rrrthree{Before continuing on from this point, we stress again that the FIRE-2
simulation suite consists of a relatively small number of initial conditions,
and that the isolated dwarfs in the simulation suite are biased towards conditions
that are more isolated than the typical field dwarf.
Due to the small number of initial conditions and the selection of
these initial conditions, it is key to remember that the FIRE-2 
dwarfs should not be treated as a dwarf population study.
Furthermore, of these
relatively small number of initial conditions, the FIRE-2 suite produces a large number
of \ndd{} dwarfs that are inconsistent with average observed dwarf galaxies 
in both their stellar structure and gas content. 
We thus emphasize
that our goal is not to validate the ability of the FIRE-2 simulations
to reproduce the overall
dwarf population, but to demonstrate that the in-situ mode of dwarf stellar halo 
creation that has been suggested from observational results can indeed be naturally
produced by contemporary cosmological hydrodynamical simulations.}

\subsection{The \NDD{} Dwarfs}\label{s:discussion:triaxial}
Although the FIRE\rrrthree{-2} simulations succeed in producing dwarfs with a disk-halo
structure that well-reflects the observations, it must be noted that these
disky dwarfs are not the dominant mode of dwarf stellar structure in the
FIRE\rrrthree{-2} simulations. Indeed, most of the dwarfs in the sample have significantly
lower values of B/A than observed dwarfs, and do not host a 
clear young stellar disk. To better understand this discrepancy, we 
suggest three possibilities below.

\fcr{First, the production of \ndd{} dwarfs at $z=0$ could stem from the
implemented physics of the FIRE-2 simulations.} 
That the \ndd{} dwarfs tend to be gas-poor
with a bursty recent SFH (see \autoref{f:burstgas}) suggests that the
star formation
feedback prescription in FIRE\rrrthree{-2} may be preventing the formation of both 
stellar and gaseous disks in these \ndd{} dwarfs \citep[see also][]{elbadry2018quataert}. 
\fcr{It is also notable that observed high redshift dwarfs lack a well-defined disk; 
\cite{zhang2019} found that high-z, high-mass dwarfs
are characterized by triaxial prolate ellipsoids at high redshift ($z\gtrsim1$) and
transition to thick-disk structures at low-z \citep[see also][for a similar transition in simulated dwarfs]{ceverino2015}. 
It is thus possible that
the FIRE-2 dwarfs are failing to undergo this transition at the appropriate cosmic time.}
At the same time,
the existence of the disky dwarfs, which also host in-situ stellar halos,
indicates that it is possible to form disk-halo structures with this
feedback prescription. 
\fcr{Interestingly, \cite{elbadry2018bradford} also showed that the low-mass FIRE\rrrthree{-2} galaxies
tend to be overly dispersion-supported due to inefficient accretion of high angular 
momentum gas, 
with the same trend that rotation-supported gas disks
occupy galaxies with less bursty recent star formation histories. 
This interplay between the
gaseous structure, stellar structure, and recent
star formation history of the FIRE\rrrthree{-2} dwarfs point to a link between 
the way in which star formation proceeds in these dwarfs and
their ability to form a young stellar disk.}
\rrr{Indeed, the idea that star formation feedback can significantly influence 
dwarf structure is well-established in regards to both the baryonic morphology 
\citep[see, e.g.][]{governato2007, smith2019, smith2020} and in the dark matter halo 
\citep[the core-cusp problem; see, e.g.][]{penarrubia2012,pontzen2012}.}
Indeed, as the most easily observable leg of this set, the 
stellar structure of galaxies in the dwarf mass regime could prove
to be a powerful tool to constrain the feedback prescriptions in conjunction
with requirements for cored/cuspy dark matter profiles.

Second, it is possible that observations are missing a significant population
of \ndd{} dwarfs due primarily to surface brightness sensitivity limits. 
We find that the \ndd{} dwarfs are not 
systematically lower in stellar surface density (within the stellar half-light radius) than the
disky dwarfs. \fcr{However, there is evidence that
the FIRE\rrrthree{-2} dwarfs with more bursty star formation also have lower surface brightnesses
\citep{chan2018}. It is thus plausible that observed samples preferentially include
disky dwarfs.}

Explaining the lack of observational analogues to \ndd{} dwarfs in FIRE\rrrthree{-2} as being due to observational incompleteness
would also necessitate the existence
of a substantial and likely dominant population of
low surface brightness dwarfs at 
$10^8\lesssim M_\star/M_\odot\lesssim10^{10}$. A forthcoming probe of the distance and 
stellar mass distribution of a sample of low surface brightness galaxies
will help to directly address this question (Greco et al., in preparation).
However, observational studies of the intrinsic shapes of low surface
brightness galaxies \citep{kadofong2021}
and ultra diffuse galaxies in clusters \citep{burkert2017,rong2019} 
indicate that although LSBGs are relatively round, 
they do not show significantly lower values of B/A than spectroscopic dwarf samples, 
as is the case for the \ndd{} dwarfs. 
We thus find it unlikely that the \ndd{} dwarfs represent a dominant
low surface brightness population that are undetected in current 
generation surveys. \fcr{Similarly, because the \ndd{} galaxies lack a disk at all
stellar age bins considered, we find it unlikely that the difference in observed and
simulated intrinsic shape distributions originates from a difference between light-weighted
differential and age-binned, mass-weighted cumulative shapes.}
\fcr{We note that some of the \ndd{} dwarfs are consistent 
with previous inferred shapes for cluster ultra diffuse galaxies \citep{rong2019}\footnote{
Though the results of \cite{burkert2017} appear to be in good agreement with the
\ndd{} dwarf population, we caution that \cite{burkert2017} did not allow for triaxiality in
their model, and their results are therefore constrained to either B/A=C/A or B/A=1.}. It is of
interest, but beyond the scope of this work, to ask whether these galaxies differ from the
general set of \ndd{} dwarfs.}

A last possibility is that, 
because the FIRE\rrrthree{-2} suite represents a relatively
small number of dwarf galaxies, \ndd{} dwarfs may be a smaller proportion of the total dwarf population than is reflected in the FIRE\rrrthree{-2} 
suite. 
Part of this discrepancy could be due to bias in the dwarfs chosen for the suite. Galaxies in FIRE\rrrthree{-2} zooms are selected to be isolated, and to have small Lagrangian regions, and so it is possible that there is some systematic bias in the sample that causes them to be more \ndd{}. 
However, observational results do not indicate that non-isolated dwarfs are more \ndd{}
\citep{kadofong2020}; further investigation into the behavior of the FIRE\rrrthree{-2} zoom suite as a 
function of environment would be enlightening, but is outside the scope of this work.
To place a limit on the sensitivity of the observational methods of
\cite{kadofong2020} to a morphologically distinct \ndd{} population, we construct a toy model
wherein we simulate a mock galaxy population composed of a disky and \ndd{} population 
and rerun our observational analysis, as detailed in \autoref{s:toymodel}. \fcr{We find that the hypothetical presence of a small minority ($\lesssim 10\%$) of galaxies with \ndd{} shapes would not affect the 
observationally inferred intrinsic shape parameters. However, the 
fraction of \ndd{} galaxies in FIRE-2 greatly exceeds this 10\% 
threshold.}
Thus, we find
that the over-representation of \ndd{} dwarfs in the simulated sample is unlikely to be explained \fcr{by observational uncertainties or}
by Poisson fluctuations ($\rm N_{\rm \ndd{}} = 10 \pm 3$).


\section{Conclusions}
In this work, we have shown that the FIRE-2 simulation suite is capable
of reproducing the young, thick stellar disk and
old, round stellar halo observed to be the dominant structural configuration 
of high-mass ($\rm 10^{8.5}<M_\star/M_\odot<10^{9.6}$) dwarf galaxies. 
These dwarf halos bear an observational
resemblance to the accreted stellar halos of higher mass galaxies, 
but as we show in \autoref{s:results:origin}, the dwarf stellar halos in FIRE\rrrthree{-2} 
are built up by the migration of in-situ stars. The FIRE\rrrthree{-2} galaxies
that succeed in forming a disk-stellar halo system also succeed in reproducing
a number of characteristics of observed dwarfs: their stellar halos
are more flattened (lower $C/A$) than the stellar halos of the higher mass
galaxies in FIRE\rrrthree{-2} (see \autoref{f:baca}), appear ubiquitously around dwarfs
that form disks, and have HI gas fractions in
good agreement with observed galaxies (see \autoref{f:bradford}). These
 dwarfs demonstrate that dwarf stellar halo
formation via the heating of disk stars is able to reproduce 
several properties of observed dwarf galaxies -- 
such an in-situ pathway appears necessary to explain the commonality of
dwarf stellar halos in observational samples. 

However, the majority of the dwarfs in FIRE\rrrthree{-2} are not disk-halo systems. 
The rest of the dwarfs in this work are instead ``\ndd{}'' systems, 
here meaning that the dwarf is significantly \ndd{} in
shape ($A\not\sim B\not\sim C$), without a clear disk-halo transition 
(see \autoref{f:baca}). We find that these \ndd{} dwarfs tend to be depleted in
HI for their stellar mass \rrrthree{or} have highly bursty recent star formation 
histories compared to the disky dwarfs. These differences may suggest that
the galaxies are unable to form a disk due to over-vigorous star formation or 
star formation feedback. 

\rrrthree{Given the presence of this non-disky population of dwarfs, we 
do not claim here that the in-situ method of dwarf stellar halo assembly
seen in the disky FIRE\rrrthree{-2} dwarfs is the definitive method by which 
stellar halos form around low mass galaxies in the real Universe. However, we have demonstrated in this work that 
when in-situ stellar outskirts are formed in FIRE-2, they
succeed in reproducing many of the structural properties of
observed dwarfs.
 Indeed, both} 
classes of FIRE\rrrthree{-2} dwarfs demonstrate that intrinsic shapes are a useful tool to understand whether simulations are able to 
produce realistic galaxy structures. This is particularly important for
low-mass galaxies where the details of the star formation history, 
star formation feedback, and gas accretion
are expected to play a significant role in shaping the overall stellar 
content and structure (both due to the shallower potential well and the lower 
fractional contribution by accreted stars).

Future observational works targeting the low surface brightness universe
will be able to determine whether there 
exists a counterpart to the \ndd{} dwarfs in FIRE\rrrthree{-2} that has been 
out of reach of previous observational studies (Greco et al., in preparation).
Indeed, a great deal of theoretical and observational work remains 
necessary to understand the population of low mass galaxies beyond the
Local Universe; among other open questions,
the details of their star formation histories \citep{chan2015}, 
their capacity to self-quench \citep{geha2012, dickey2020}, and the emergence 
of ultra diffuse galaxies \citep{chan2018,greco2018, wright2020, kadofong2021} remain
areas of active study for both observational and theoretical efforts.
All three of these questions are tied intrinsically to the morphology
(and more generally, the structure) of the dwarf population -- 
future observational studies of the intrinsic shape of the dwarf population
and other structural parameters will thus help to shed light on a number
of open questions in dwarf evolution and stellar assembly.

\acknowledgements 
EKF thanks Alexander Gurvich for thoughtful comments and discussion that improved the quality of this manuscript. \rrrthree{The authors
thank the anonymous referee for their helpful comments that have
improved the content of this work.}
JEG gratefully acknowledges support from NSF grant AST-1907723.
RES acknowledges support from NASA grant 19-ATP19-0068 and HST-AR-15809 from the Space Telescope Science Institute (STScI), which is operated by AURA, Inc., under NASA contract NAS5-26555. TKC was supported by Science and Technology Facilities Council (STFC) astronomy consolidated grant ST/P000541/1 and ST/T000244/1. AW received support from NASA through ATP grants 80NSSC18K1097 and 80NSSC20K0513; HST grants GO-14734, AR-15057, AR-15809, and GO-15902 from STScI; a Scialog Award from the Heising-Simons Foundation; and a Hellman Fellowship.
CAFG was supported by NSF through grants AST-1715216 and CAREER award AST-1652522; by NASA through grant 17-ATP17-0067; by STScI through grant HST-AR-16124.001-A; and by the Research Corporation for Science Advancement through a Cottrell Scholar Award and a Scialog Award.
Support for PFH was provided by NSF Research Grants 1911233 \&\ 20009234, NSF CAREER grant 1455342, NASA grants 80NSSC18K0562, HST-AR-15800.001-A. Numerical calculations were run on the Caltech compute cluster ``Wheeler,'' allocations FTA-Hopkins/AST20016 supported by the NSF and TACC, and NASA HEC SMD-16-7592. MBK acknowledges support from NSF CAREER award AST-1752913, NSF grant AST-1910346, NASA grant NNX17AG29G, and HST-AR-15006, HST-AR-15809, HST-GO-15658, HST-GO-15901, HST-GO-15902, HST-AR-16159, and HST-GO-16226 from STScI.

Simulations used in this work were run using XSEDE supported by NSF grant ACI-1548562, Blue Waters via allocation PRAC NSF.1713353 supported by the NSF, and NASA HEC Program through the NAS Division at Ames Research Center.

The authors also would like to thank the Flatiron Institute Scientific Computing Core for providing computing resources that made this research possible, and especially for their hard work facilitating remote work during the pandemic. Analysis for this paper was carried out on the Flatiron Institute's computing cluster \texttt{rusty}, which is supported by the Simons Foundation.

\software{GizmoAnalysis \citep{wetzel2020}, 
matplotlib \citep{Hunter:2007}, 
SciPy \citep{jones_scipy_2001}, 
the IPython package \citep{PER-GRA:2007}, and  
NumPy \citep{van2011numpy}}

\bibliography{FIRE-shapes.bib,shapes-RES.bib}

\begin{thebibliography}{}
\providecommand\natexlab[1]{#1}
\providecommand\JournalTitle[1]{#1}

\bibitem[{jon(2001)}]{jones_scipy_2001}
 2001, {SciPy}: Open source scientific tools for Python

\bibitem[{{Agertz} {et~al.}(2021){Agertz}, {Renaud}, {Feltzing}, {Read},
  {Ryde}, {Andersson}, {Rey}, {Bensby}, \& {Feuillet}}]{agertz2021}
{Agertz}, O., {Renaud}, F., {Feltzing}, S., {et~al.} 2021,
  \href{http://dx.doi.org/10.1093/mnras/stab322}{\JournalTitle{\mnras}, 503,
  5826}

\bibitem[{{Allgood} {et~al.}(2006){Allgood}, {Flores}, {Primack}, {Kravtsov},
  {Wechsler}, {Faltenbacher}, \& {Bullock}}]{allgood2006}
{Allgood}, B., {Flores}, R.~A., {Primack}, J.~R., {et~al.} 2006,
  \href{http://dx.doi.org/10.1111/j.1365-2966.2006.10094.x}{\JournalTitle{\mnras},
  367, 1781}

\bibitem[{{Angl{\'e}s-Alc{\'a}zar} {et~al.}(2017){Angl{\'e}s-Alc{\'a}zar},
  {Faucher-Gigu{\`e}re}, {Kere{\v{s}}}, {Hopkins}, {Quataert}, \&
  {Murray}}]{anglesalcazar2017}
{Angl{\'e}s-Alc{\'a}zar}, D., {Faucher-Gigu{\`e}re}, C.-A., {Kere{\v{s}}}, D.,
  {et~al.} 2017,
  \href{http://dx.doi.org/10.1093/mnras/stx1517}{\JournalTitle{\mnras}, 470,
  4698}

\bibitem[{{Aparicio} \& {Tikhonov}(2000)}]{aparicio2000}
{Aparicio}, A., \& {Tikhonov}, N. 2000,
  \href{http://dx.doi.org/10.1086/301360}{\JournalTitle{\aj}, 119, 2183}

\bibitem[{{Aparicio} {et~al.}(2000){Aparicio}, {Tikhonov}, \&
  {Karachentsev}}]{aparicio2000a}
{Aparicio}, A., {Tikhonov}, N., \& {Karachentsev}, I. 2000,
  \href{http://dx.doi.org/10.1086/301157}{\JournalTitle{\aj}, 119, 177}

\bibitem[{{Belokurov} {et~al.}(2020){Belokurov}, {Sanders}, {Fattahi}, {Smith},
  {Deason}, {Evans}, \& {Grand}}]{belokurov2020}
{Belokurov}, V., {Sanders}, J.~L., {Fattahi}, A., {et~al.} 2020,
  \href{http://dx.doi.org/10.1093/mnras/staa876}{\JournalTitle{\mnras}, 494,
  3880}

\bibitem[{{Benincasa} {et~al.}(2020){Benincasa}, {Loebman}, {Wetzel},
  {Hopkins}, {Murray}, {Bellardini}, {Faucher-Gigu{\`e}re}, {Guszejnov}, \&
  {Orr}}]{2020MNRAS.497.3993B}
{Benincasa}, S.~M., {Loebman}, S.~R., {Wetzel}, A., {et~al.} 2020,
  \href{http://dx.doi.org/10.1093/mnras/staa2116}{\JournalTitle{\mnras}, 497,
  3993}

\bibitem[{{Bernard} {et~al.}(2007){Bernard}, {Aparicio}, {Gallart},
  {Padilla-Torres}, \& {Panniello}}]{bernard2007}
{Bernard}, E.~J., {Aparicio}, A., {Gallart}, C., {Padilla-Torres}, C.~P., \&
  {Panniello}, M. 2007,
  \href{http://dx.doi.org/10.1086/520805}{\JournalTitle{\aj}, 134, 1124}

\bibitem[{{Bovill} \& {Ricotti}(2009)}]{2009ApJ...693.1859B}
{Bovill}, M.~S., \& {Ricotti}, M. 2009,
  \href{http://dx.doi.org/10.1088/0004-637X/693/2/1859}{\JournalTitle{\apj},
  693, 1859}

\bibitem[{{Bovy} {et~al.}(2012){Bovy}, {Rix}, \& {Hogg}}]{bovy2012}
{Bovy}, J., {Rix}, H.-W., \& {Hogg}, D.~W. 2012,
  \href{http://dx.doi.org/10.1088/0004-637X/751/2/131}{\JournalTitle{\apj},
  751, 131}

\bibitem[{{Boylan-Kolchin} {et~al.}(2015){Boylan-Kolchin}, {Weisz}, {Johnson},
  {Bullock}, {Conroy}, \& {Fitts}}]{2015MNRAS.453.1503B}
{Boylan-Kolchin}, M., {Weisz}, D.~R., {Johnson}, B.~D., {et~al.} 2015,
  \href{http://dx.doi.org/10.1093/mnras/stv1736}{\JournalTitle{\mnras}, 453,
  1503}

\bibitem[{{Bradford} {et~al.}(2015){Bradford}, {Geha}, \&
  {Blanton}}]{bradford2015}
{Bradford}, J.~D., {Geha}, M.~C., \& {Blanton}, M.~R. 2015,
  \href{http://dx.doi.org/10.1088/0004-637X/809/2/146}{\JournalTitle{\apj},
  809, 146}

\bibitem[{{Brook} {et~al.}(2014){Brook}, {Di Cintio}, {Knebe}, {Gottl{\"o}ber},
  {Hoffman}, {Yepes}, \& {Garrison-Kimmel}}]{brook2014}
{Brook}, C.~B., {Di Cintio}, A., {Knebe}, A., {et~al.} 2014,
  \href{http://dx.doi.org/10.1088/2041-8205/784/1/L14}{\JournalTitle{\apjl},
  784, L14}

\bibitem[{{Brooks} \& {Zolotov}(2014)}]{brooks2014}
{Brooks}, A.~M., \& {Zolotov}, A. 2014,
  \href{http://dx.doi.org/10.1088/0004-637X/786/2/87}{\JournalTitle{\apj}, 786,
  87}

\bibitem[{{Bryan} \& {Norman}(1998)}]{bn1998}
{Bryan}, G.~L., \& {Norman}, M.~L. 1998,
  \href{http://dx.doi.org/10.1086/305262}{\JournalTitle{\apj}, 495, 80}

\bibitem[{{Bullock} \& {Johnston}(2005)}]{bullock2005}
{Bullock}, J.~S., \& {Johnston}, K.~V. 2005,
  \href{http://dx.doi.org/10.1086/497422}{\JournalTitle{\apj}, 635, 931}

\bibitem[{{Burkert}(2017)}]{burkert2017}
{Burkert}, A. 2017,
  \href{http://dx.doi.org/10.3847/1538-4357/aa671c}{\JournalTitle{\apj}, 838,
  93}

\bibitem[{{Carlsten} {et~al.}(2021){Carlsten}, {Greene}, {Greco}, {Beaton}, \&
  {Kado-Fong}}]{carlsten2021}
{Carlsten}, S.~G., {Greene}, J.~E., {Greco}, J.~P., {Beaton}, R.~L., \&
  {Kado-Fong}, E. 2021, \JournalTitle{arXiv e-prints}, arXiv:2105.03435

\bibitem[{{Ceverino} {et~al.}(2015){Ceverino}, {Primack}, \&
  {Dekel}}]{ceverino2015}
{Ceverino}, D., {Primack}, J., \& {Dekel}, A. 2015,
  \href{http://dx.doi.org/10.1093/mnras/stv1603}{\JournalTitle{\mnras}, 453,
  408}

\bibitem[{{Chan} {et~al.}(2019){Chan}, {Kere{\v{s}}}, {Hopkins}, {Quataert},
  {Su}, {Hayward}, \& {Faucher-Gigu{\`e}re}}]{chan2019}
{Chan}, T.~K., {Kere{\v{s}}}, D., {Hopkins}, P.~F., {et~al.} 2019,
  \href{http://dx.doi.org/10.1093/mnras/stz1895}{\JournalTitle{\mnras}, 488,
  3716}

\bibitem[{{Chan} {et~al.}(2015){Chan}, {Kere{\v{s}}}, {O{\~n}orbe}, {Hopkins},
  {Muratov}, {Faucher-Gigu{\`e}re}, \& {Quataert}}]{chan2015}
{Chan}, T.~K., {Kere{\v{s}}}, D., {O{\~n}orbe}, J., {et~al.} 2015,
  \href{http://dx.doi.org/10.1093/mnras/stv2165}{\JournalTitle{\mnras}, 454,
  2981}

\bibitem[{{Chan} {et~al.}(2018){Chan}, {Kere{\v{s}}}, {Wetzel}, {Hopkins},
  {Faucher-Gigu{\`e}re}, {El-Badry}, {Garrison-Kimmel}, \&
  {Boylan-Kolchin}}]{chan2018}
{Chan}, T.~K., {Kere{\v{s}}}, D., {Wetzel}, A., {et~al.} 2018,
  \href{http://dx.doi.org/10.1093/mnras/sty1153}{\JournalTitle{\mnras}, 478,
  906}

\bibitem[{{Conroy} \& {Gunn}(2010)}]{conroy2010}
{Conroy}, C., \& {Gunn}, J.~E. 2010,
  \href{http://dx.doi.org/10.1088/0004-637X/712/2/833}{\JournalTitle{\apj},
  712, 833}

\bibitem[{{Conroy} {et~al.}(2009){Conroy}, {Gunn}, \& {White}}]{conroy2009}
{Conroy}, C., {Gunn}, J.~E., \& {White}, M. 2009,
  \href{http://dx.doi.org/10.1088/0004-637X/699/1/486}{\JournalTitle{\apj},
  699, 486}

\bibitem[{{Das} \& {Binney}(2016)}]{das2016}
{Das}, P., \& {Binney}, J. 2016,
  \href{http://dx.doi.org/10.1093/mnras/stw744}{\JournalTitle{\mnras}, 460,
  1725}

\bibitem[{{Dashyan} \& {Dubois}(2020)}]{dashyan2020}
{Dashyan}, G., \& {Dubois}, Y. 2020,
  \href{http://dx.doi.org/10.1051/0004-6361/201936339}{\JournalTitle{\aap},
  638, A123}

\bibitem[{{Demers} {et~al.}(2006){Demers}, {Battinelli}, \&
  {Artigau}}]{demers2006}
{Demers}, S., {Battinelli}, P., \& {Artigau}, E. 2006,
  \href{http://dx.doi.org/10.1051/0004-6361:20065507}{\JournalTitle{\aap}, 456,
  905}

\bibitem[{{Dickey} {et~al.}(2020){Dickey}, {Starkenburg}, {Geha}, {Hahn},
  {Angl{\'e}s-Alc{\'a}zar}, {Choi}, {Dav{\'e}}, {Genel}, {Iyer}, {Maller},
  {Mandelker}, {Somerville}, \& {Yung}}]{dickey2020}
{Dickey}, C.~M., {Starkenburg}, T.~K., {Geha}, M., {et~al.} 2020,
  \JournalTitle{arXiv e-prints}, arXiv:2010.01132

\bibitem[{{El-Badry} {et~al.}(2016){El-Badry}, {Wetzel}, {Geha}, {Hopkins},
  {Kere{\v{s}}}, {Chan}, \& {Faucher-Gigu{\`e}re}}]{elbadry2016}
{El-Badry}, K., {Wetzel}, A., {Geha}, M., {et~al.} 2016,
  \href{http://dx.doi.org/10.3847/0004-637X/820/2/131}{\JournalTitle{\apj},
  820, 131}

\bibitem[{{El-Badry} {et~al.}(2018{\natexlab{a}}){El-Badry}, {Bradford},
  {Quataert}, {Geha}, {Boylan-Kolchin}, {Weisz}, {Wetzel}, {Hopkins}, {Chan},
  {Fitts}, {Kere{\v{s}}}, \& {Faucher-Gigu{\`e}re}}]{elbadry2018bradford}
{El-Badry}, K., {Bradford}, J., {Quataert}, E., {et~al.} 2018{\natexlab{a}},
  \href{http://dx.doi.org/10.1093/mnras/sty730}{\JournalTitle{\mnras}, 477,
  1536}

\bibitem[{{El-Badry} {et~al.}(2018{\natexlab{b}}){El-Badry}, {Quataert},
  {Wetzel}, {Hopkins}, {Weisz}, {Chan}, {Fitts}, {Boylan-Kolchin},
  {Kere{\v{s}}}, {Faucher-Gigu{\`e}re}, \&
  {Garrison-Kimmel}}]{elbadry2018quataert}
{El-Badry}, K., {Quataert}, E., {Wetzel}, A., {et~al.} 2018{\natexlab{b}},
  \href{http://dx.doi.org/10.1093/mnras/stx2482}{\JournalTitle{\mnras}, 473,
  1930}

\bibitem[{{El-Badry} {et~al.}(2018{\natexlab{c}}){El-Badry}, {Quataert},
  {Wetzel}, {Hopkins}, {Weisz}, {Chan}, {Fitts}, {Boylan-Kolchin},
  {Kere{\v{s}}}, {Faucher-Gigu{\`e}re}, \& {Garrison-Kimmel}}]{elbadry2018}
---. 2018{\natexlab{c}},
  \href{http://dx.doi.org/10.1093/mnras/stx2482}{\JournalTitle{\mnras}, 473,
  1930}

\bibitem[{{Emami} {et~al.}(2021){Emami}, {Siana}, {El-Badry}, {Cook}, {Ma},
  {Weisz}, {Gharibshah}, {Alaee}, {Scarlata}, \& {Skillman}}]{emami2021}
{Emami}, N., {Siana}, B., {El-Badry}, K., {et~al.} 2021, \JournalTitle{arXiv
  e-prints}, arXiv:2108.08857

\bibitem[{{Escala} {et~al.}(2018){Escala}, {Wetzel}, {Kirby}, {Hopkins}, {Ma},
  {Wheeler}, {Kere{\v{s}}}, {Faucher-Gigu{\`e}re}, \& {Quataert}}]{escala}
{Escala}, I., {Wetzel}, A., {Kirby}, E.~N., {et~al.} 2018,
  \href{http://dx.doi.org/10.1093/mnras/stx2858}{\JournalTitle{\mnras}, 474,
  2194}

\bibitem[{{Faucher-Gigu{\`e}re}(2018)}]{fauchergiguere2018}
{Faucher-Gigu{\`e}re}, C.-A. 2018,
  \href{http://dx.doi.org/10.1093/mnras/stx2595}{\JournalTitle{\mnras}, 473,
  3717}

\bibitem[{{Faucher-Gigu{\`e}re} {et~al.}(2009){Faucher-Gigu{\`e}re}, {Lidz},
  {Zaldarriaga}, \& {Hernquist}}]{FaucherGiguere2009}
{Faucher-Gigu{\`e}re}, C.-A., {Lidz}, A., {Zaldarriaga}, M., \& {Hernquist}, L.
  2009,
  \href{http://dx.doi.org/10.1088/0004-637X/703/2/1416}{\JournalTitle{\apj},
  703, 1416}

\bibitem[{{Fitts} {et~al.}(2017){Fitts}, {Boylan-Kolchin}, {Elbert}, {Bullock},
  {Hopkins}, {O{\~n}orbe}, {Wetzel}, {Wheeler}, {Faucher-Gigu{\`e}re},
  {Kere{\v{s}}}, {Skillman}, \& {Weisz}}]{2017MNRAS.471.3547F}
{Fitts}, A., {Boylan-Kolchin}, M., {Elbert}, O.~D., {et~al.} 2017,
  \href{http://dx.doi.org/10.1093/mnras/stx1757}{\JournalTitle{\mnras}, 471,
  3547}

\bibitem[{{Fitts} {et~al.}(2018){Fitts}, {Boylan-Kolchin}, {Bullock}, {Weisz},
  {El-Badry}, {Wheeler}, {Faucher-Gigu{\`e}re}, {Quataert}, {Hopkins},
  {Kere{\v{s}}}, {Wetzel}, \& {Hayward}}]{fitts2018}
{Fitts}, A., {Boylan-Kolchin}, M., {Bullock}, J.~S., {et~al.} 2018,
  \href{http://dx.doi.org/10.1093/mnras/sty1488}{\JournalTitle{\mnras}, 479,
  319}

\bibitem[{{Flores Vel{\'a}zquez} {et~al.}(2021){Flores Vel{\'a}zquez},
  {Gurvich}, {Faucher-Gigu{\`e}re}, {Bullock}, {Starkenburg}, {Moreno},
  {Lazar}, {Mercado}, {Stern}, {Sparre}, {Hayward}, {Wetzel}, \&
  {El-Badry}}]{florevelazquez2021}
{Flores Vel{\'a}zquez}, J.~A., {Gurvich}, A.~B., {Faucher-Gigu{\`e}re}, C.-A.,
  {et~al.} 2021,
  \href{http://dx.doi.org/10.1093/mnras/staa3893}{\JournalTitle{\mnras}, 501,
  4812}

\bibitem[{{Foreman-Mackey} {et~al.}(2013){Foreman-Mackey}, {Hogg}, {Lang}, \&
  {Goodman}}]{foremanmackey2013}
{Foreman-Mackey}, D., {Hogg}, D.~W., {Lang}, D., \& {Goodman}, J. 2013,
  \href{http://dx.doi.org/10.1086/670067}{\JournalTitle{\pasp}, 125, 306}

\bibitem[{{Garrison-Kimmel} {et~al.}(2019){Garrison-Kimmel}, {Wetzel},
  {Hopkins}, {Sanderson}, {El-Badry}, {Graus}, {Chan}, {Feldmann},
  {Boylan-Kolchin}, {Hayward}, {Bullock}, {Fitts}, {Samuel}, {Wheeler},
  {Kere{\v{s}}}, \& {Faucher-Gigu{\`e}re}}]{2019MNRAS.489.4574G}
{Garrison-Kimmel}, S., {Wetzel}, A., {Hopkins}, P.~F., {et~al.} 2019,
  \href{http://dx.doi.org/10.1093/mnras/stz2507}{\JournalTitle{\mnras}, 489,
  4574}

\bibitem[{{Geha} {et~al.}(2012){Geha}, {Blanton}, {Yan}, \&
  {Tinker}}]{geha2012}
{Geha}, M., {Blanton}, M.~R., {Yan}, R., \& {Tinker}, J.~L. 2012,
  \href{http://dx.doi.org/10.1088/0004-637X/757/1/85}{\JournalTitle{\apj}, 757,
  85}

\bibitem[{{Governato} {et~al.}(2007){Governato}, {Willman}, {Mayer}, {Brooks},
  {Stinson}, {Valenzuela}, {Wadsley}, \& {Quinn}}]{governato2007}
{Governato}, F., {Willman}, B., {Mayer}, L., {et~al.} 2007,
  \href{http://dx.doi.org/10.1111/j.1365-2966.2006.11266.x}{\JournalTitle{\mnras},
  374, 1479}

\bibitem[{{Graus} {et~al.}(2019{\natexlab{a}}){Graus}, {Bullock}, {Fitts},
  {Cooper}, {Boylan-Kolchin}, {Weisz}, {Wetzel}, {Feldmann},
  {Faucher-Gigu{\`e}re}, {Quataert}, {Hopkins}, \&
  {Kere{\v{s}}}}]{2019MNRAS.490.1186G}
{Graus}, A.~S., {Bullock}, J.~S., {Fitts}, A., {et~al.} 2019{\natexlab{a}},
  \href{http://dx.doi.org/10.1093/mnras/stz2649}{\JournalTitle{\mnras}, 490,
  1186}

\bibitem[{{Graus} {et~al.}(2019{\natexlab{b}}){Graus}, {Bullock}, {Fitts},
  {Cooper}, {Boylan-Kolchin}, {Weisz}, {Wetzel}, {Feldmann},
  {Faucher-Gigu{\`e}re}, {Quataert}, {Hopkins}, \& {Kere{\v{s}}}}]{graus2019}
---. 2019{\natexlab{b}},
  \href{http://dx.doi.org/10.1093/mnras/stz2649}{\JournalTitle{\mnras}, 490,
  1186}

\bibitem[{{Greco} {et~al.}(2018){Greco}, {Greene}, {Strauss}, {Macarthur},
  {Flowers}, {Goulding}, {Huang}, {Kim}, {Komiyama}, {Leauthaud}, {Leisman},
  {Lupton}, {Sif{\'o}n}, \& {Wang}}]{greco2018}
{Greco}, J.~P., {Greene}, J.~E., {Strauss}, M.~A., {et~al.} 2018,
  \href{http://dx.doi.org/10.3847/1538-4357/aab842}{\JournalTitle{\apj}, 857,
  104}

\bibitem[{{Hargis} {et~al.}(2020){Hargis}, {Albers}, {Crnojevi{\'c}}, {Sand },
  {Weisz}, {Carlin}, {Spekkens}, {Willman}, {Peter}, {Grillmair}, \&
  {Dolphin}}]{hargis2020}
{Hargis}, J.~R., {Albers}, S., {Crnojevi{\'c}}, D., {et~al.} 2020,
  \href{http://dx.doi.org/10.3847/1538-4357/ab58d2}{\JournalTitle{\apj}, 888,
  31}

\bibitem[{{Hidalgo} {et~al.}(2003){Hidalgo}, {Mar{\'\i}n-Franch}, \&
  {Aparicio}}]{hidalgo2003}
{Hidalgo}, S.~L., {Mar{\'\i}n-Franch}, A., \& {Aparicio}, A. 2003,
  \href{http://dx.doi.org/10.1086/367779}{\JournalTitle{\aj}, 125, 1247}

\bibitem[{{Hopkins}(2015)}]{hopkins2015}
{Hopkins}, P.~F. 2015,
  \href{http://dx.doi.org/10.1093/mnras/stv195}{\JournalTitle{\mnras}, 450, 53}

\bibitem[{{Hopkins}(2016)}]{hopkins2016}
---. 2016,
  \href{http://dx.doi.org/10.1093/mnras/stw1578}{\JournalTitle{\mnras}, 462,
  576}

\bibitem[{{Hopkins}(2017)}]{hopkins2017}
---. 2017,
  \href{http://dx.doi.org/10.1093/mnras/stw3306}{\JournalTitle{\mnras}, 466,
  3387}

\bibitem[{{Hopkins} {et~al.}(2020{\natexlab{a}}){Hopkins}, {Grudi{\'c}},
  {Wetzel}, {Kere{\v{s}}}, {Faucher-Gigu{\`e}re}, {Ma}, {Murray}, \&
  {Butcher}}]{2020MNRAS.491.3702H}
{Hopkins}, P.~F., {Grudi{\'c}}, M.~Y., {Wetzel}, A., {et~al.}
  2020{\natexlab{a}},
  \href{http://dx.doi.org/10.1093/mnras/stz3129}{\JournalTitle{\mnras}, 491,
  3702}

\bibitem[{{Hopkins} {et~al.}(2013){Hopkins}, {Narayanan}, \&
  {Murray}}]{Hopkins2013sf_criteria}
{Hopkins}, P.~F., {Narayanan}, D., \& {Murray}, N. 2013,
  \href{http://dx.doi.org/10.1093/mnras/stt723}{\JournalTitle{\mnras}, 432,
  2647}

\bibitem[{{Hopkins} \& {Raives}(2016)}]{hr2016}
{Hopkins}, P.~F., \& {Raives}, M.~J. 2016,
  \href{http://dx.doi.org/10.1093/mnras/stv2180}{\JournalTitle{\mnras}, 455,
  51}

\bibitem[{{Hopkins} {et~al.}(2018){Hopkins}, {Wetzel}, {Kere{\v{s}}},
  {Faucher-Gigu{\`e}re}, {Quataert}, {Boylan-Kolchin}, {Murray}, {Hayward},
  {Garrison-Kimmel}, {Hummels}, {Feldmann}, {Torrey}, {Ma},
  {Angl{\'e}s-Alc{\'a}zar}, {Su}, {Orr}, {Schmitz}, {Escala}, {Sanderson},
  {Grudi{\'c}}, {Hafen}, {Kim}, {Fitts}, {Bullock}, {Wheeler}, {Chan},
  {Elbert}, \& {Narayanan}}]{hopkins2018_main}
{Hopkins}, P.~F., {Wetzel}, A., {Kere{\v{s}}}, D., {et~al.} 2018,
  \href{http://dx.doi.org/10.1093/mnras/sty1690}{\JournalTitle{\mnras}, 480,
  800}

\bibitem[{{Hopkins} {et~al.}(2020{\natexlab{b}}){Hopkins}, {Chan},
  {Garrison-Kimmel}, {Ji}, {Su}, {Hummels}, {Kere{\v{s}}}, {Quataert}, \&
  {Faucher-Gigu{\`e}re}}]{hopkins2020}
{Hopkins}, P.~F., {Chan}, T.~K., {Garrison-Kimmel}, S., {et~al.}
  2020{\natexlab{b}},
  \href{http://dx.doi.org/10.1093/mnras/stz3321}{\JournalTitle{\mnras}, 492,
  3465}

\bibitem[{{Hu}(2019)}]{hu2019}
{Hu}, C.-Y. 2019,
  \href{http://dx.doi.org/10.1093/mnras/sty3252}{\JournalTitle{\mnras}, 483,
  3363}

\bibitem[{{Hu} {et~al.}(2016){Hu}, {Naab}, {Walch}, {Glover}, \&
  {Clark}}]{hu2016}
{Hu}, C.-Y., {Naab}, T., {Walch}, S., {Glover}, S. C.~O., \& {Clark}, P.~C.
  2016, \href{http://dx.doi.org/10.1093/mnras/stw544}{\JournalTitle{\mnras},
  458, 3528}

\bibitem[{Hunter(2007)}]{Hunter:2007}
Hunter, J.~D. 2007,
  \href{http://dx.doi.org/10.1109/MCSE.2007.55}{\JournalTitle{Computing in
  Science \& Engineering}, 9, 90}

\bibitem[{{Iorio} {et~al.}(2018){Iorio}, {Belokurov}, {Erkal}, {Koposov},
  {Nipoti}, \& {Fraternali}}]{iorio2018}
{Iorio}, G., {Belokurov}, V., {Erkal}, D., {et~al.} 2018,
  \href{http://dx.doi.org/10.1093/mnras/stx2819}{\JournalTitle{\mnras}, 474,
  2142}

\bibitem[{{Jahn} {et~al.}(2019){Jahn}, {Sales}, {Wetzel}, {Boylan-Kolchin},
  {Chan}, {El-Badry}, {Lazar}, \& {Bullock}}]{2019MNRAS.489.5348J}
{Jahn}, E.~D., {Sales}, L.~V., {Wetzel}, A., {et~al.} 2019,
  \href{http://dx.doi.org/10.1093/mnras/stz2457}{\JournalTitle{\mnras}, 489,
  5348}

\bibitem[{{Ji} {et~al.}(2020){Ji}, {Chan}, {Hummels}, {Hopkins}, {Stern},
  {Kere{\v{s}}}, {Quataert}, {Faucher-Gigu{\`e}re}, \& {Murray}}]{ji2020}
{Ji}, S., {Chan}, T.~K., {Hummels}, C.~B., {et~al.} 2020,
  \href{http://dx.doi.org/10.1093/mnras/staa1849}{\JournalTitle{\mnras}, 496,
  4221}

\bibitem[{{Kado-Fong} {et~al.}(2020){Kado-Fong}, {Greene}, {Huang}, {Beaton},
  {Goulding}, \& {Komiyama}}]{kadofong2020}
{Kado-Fong}, E., {Greene}, J.~E., {Huang}, S., {et~al.} 2020,
  \href{http://dx.doi.org/10.3847/1538-4357/abacc2}{\JournalTitle{\apj}, 900,
  163}

\bibitem[{{Kado-Fong} {et~al.}(2021){Kado-Fong}, {Petrescu}, {Mohammad},
  {Greco}, {Greene}, {Adams}, {Huang}, {Leisman}, {Munshi}, {Tanoglidis}, \&
  {Van Nest}}]{kadofong2021}
{Kado-Fong}, E., {Petrescu}, M., {Mohammad}, M., {et~al.} 2021,
  \JournalTitle{arXiv e-prints}, arXiv:2106.05288

\bibitem[{{Kroupa}(2001)}]{Kroupa2001}
{Kroupa}, P. 2001,
  \href{http://dx.doi.org/10.1046/j.1365-8711.2001.04022.x}{\JournalTitle{\mnras},
  322, 231}

\bibitem[{{Krumholz} \& {Gnedin}(2011)}]{Krumholz2011}
{Krumholz}, M.~R., \& {Gnedin}, N.~Y. 2011,
  \href{http://dx.doi.org/10.1088/0004-637X/729/1/36}{\JournalTitle{\apj}, 729,
  36}

\bibitem[{{Leitherer} {et~al.}(1999){Leitherer}, {Schaerer}, {Goldader},
  {Delgado}, {Robert}, {Kune}, {de Mello}, {Devost}, \&
  {Heckman}}]{leitherer1999}
{Leitherer}, C., {Schaerer}, D., {Goldader}, J.~D., {et~al.} 1999,
  \href{http://dx.doi.org/10.1086/313233}{\JournalTitle{\apjs}, 123, 3}

\bibitem[{{Lin} \& {Faber}(1983)}]{lin1983}
{Lin}, D.~N.~C., \& {Faber}, S.~M. 1983,
  \href{http://dx.doi.org/10.1086/183971}{\JournalTitle{\apjl}, 266, L21}

\bibitem[{{Ludlow} {et~al.}(2021){Ludlow}, {Fall}, {Schaye}, \&
  {Obreschkow}}]{ludlow2021}
{Ludlow}, A.~D., {Fall}, S.~M., {Schaye}, J., \& {Obreschkow}, D. 2021,
  \JournalTitle{arXiv e-prints}, arXiv:2105.03561

\bibitem[{{Maxwell} {et~al.}(2012){Maxwell}, {Wadsley}, {Couchman}, \&
  {Mashchenko}}]{maxwell2012}
{Maxwell}, A.~J., {Wadsley}, J., {Couchman}, H.~M.~P., \& {Mashchenko}, S.
  2012,
  \href{http://dx.doi.org/10.1088/2041-8205/755/2/L35}{\JournalTitle{\apjl},
  755, L35}

\bibitem[{{Minniti} \& {Zijlstra}(1996)}]{minniti1996}
{Minniti}, D., \& {Zijlstra}, A.~A. 1996,
  \href{http://dx.doi.org/10.1086/310189}{\JournalTitle{\apjl}, 467, L13}

\bibitem[{{Minniti} {et~al.}(1999){Minniti}, {Zijlstra}, \&
  {Alonso}}]{minniti1999}
{Minniti}, D., {Zijlstra}, A.~A., \& {Alonso}, M.~V. 1999,
  \href{http://dx.doi.org/10.1086/300735}{\JournalTitle{\aj}, 117, 881}

\bibitem[{{Nath Patra}(2020)}]{nathpatra2020}
{Nath Patra}, N. 2020, \JournalTitle{arXiv e-prints}, arXiv:2005.05979

\bibitem[{{Navarro} {et~al.}(1996){Navarro}, {Eke}, \& {Frenk}}]{navarro1996}
{Navarro}, J.~F., {Eke}, V.~R., \& {Frenk}, C.~S. 1996,
  \href{http://dx.doi.org/10.1093/mnras/283.3.L72}{\JournalTitle{\mnras}, 283,
  L72}

\bibitem[{{Necib} {et~al.}(2019){Necib}, {Lisanti}, {Garrison-Kimmel},
  {Wetzel}, {Sanderson}, {Hopkins}, {Faucher-Gigu{\`e}re}, \&
  {Kere{\v{s}}}}]{necib2019}
{Necib}, L., {Lisanti}, M., {Garrison-Kimmel}, S., {et~al.} 2019,
  \href{http://dx.doi.org/10.3847/1538-4357/ab3afc}{\JournalTitle{\apj}, 883,
  27}

\bibitem[{{Nidever} {et~al.}(2019{\natexlab{a}}){Nidever}, {Olsen}, {Choi}, {de
  Boer}, {Blum}, {Bell}, {Zaritsky}, {Martin}, {Saha}, {Conn}, {Besla}, {van
  der Marel}, {No{\"e}l}, {Monachesi}, {Stringfellow}, {Massana}, {Cioni},
  {Gallart}, {Monelli}, {Martinez-Delgado}, {Mu{\~n}oz}, {Majewski}, {Vivas},
  {Walker}, {Kaleida}, \& {Chu}}]{nidever2019}
{Nidever}, D.~L., {Olsen}, K., {Choi}, Y., {et~al.} 2019{\natexlab{a}},
  \href{http://dx.doi.org/10.3847/1538-4357/aafaf7}{\JournalTitle{\apj}, 874,
  118}

\bibitem[{{Nidever} {et~al.}(2019{\natexlab{b}}){Nidever}, {Olsen}, {Choi}, {de
  Boer}, {Blum}, {Bell}, {Zaritsky}, {Martin}, {Saha}, {Conn}, {Besla}, {van
  der Marel}, {No{\"e}l}, {Monachesi}, {Stringfellow}, {Massana}, {Cioni},
  {Gallart}, {Monelli}, {Martinez-Delgado}, {Mu{\~n}oz}, {Majewski}, {Vivas},
  {Walker}, {Kaleida}, \& {Chu}}]{nidever2019a}
---. 2019{\natexlab{b}},
  \href{http://dx.doi.org/10.3847/1538-4357/aafaf7}{\JournalTitle{\apj}, 874,
  118}

\bibitem[{{Padilla} \& {Strauss}(2008)}]{padilla2008}
{Padilla}, N.~D., \& {Strauss}, M.~A. 2008,
  \href{http://dx.doi.org/10.1111/j.1365-2966.2008.13480.x}{\JournalTitle{\mnras},
  388, 1321}

\bibitem[{{Park} {et~al.}(2021){Park}, {Yi}, {Peirani}, {Pichon}, {Dubois},
  {Choi}, {Devriendt}, {Kaviraj}, {Kimm}, {Kraljic}, \& {Volonteri}}]{park2021}
{Park}, M.~J., {Yi}, S.~K., {Peirani}, S., {et~al.} 2021,
  \href{http://dx.doi.org/10.3847/1538-4365/abe937}{\JournalTitle{\apjs}, 254,
  2}

\bibitem[{{Pe{\~n}arrubia} {et~al.}(2012){Pe{\~n}arrubia}, {Pontzen}, {Walker},
  \& {Koposov}}]{penarrubia2012}
{Pe{\~n}arrubia}, J., {Pontzen}, A., {Walker}, M.~G., \& {Koposov}, S.~E. 2012,
  \href{http://dx.doi.org/10.1088/2041-8205/759/2/L42}{\JournalTitle{\apjl},
  759, L42}

\bibitem[{P\'erez \& Granger(2007)}]{PER-GRA:2007}
P\'erez, F., \& Granger, B.~E. 2007,
  \href{http://dx.doi.org/10.1109/MCSE.2007.53}{\JournalTitle{Computing in
  Science and Engineering}, 9, 21}

\bibitem[{{Pillepich} {et~al.}(2018){Pillepich}, {Springel}, {Nelson}, {Genel},
  {Naiman}, {Pakmor}, {Hernquist}, {Torrey}, {Vogelsberger}, {Weinberger}, \&
  {Marinacci}}]{pillepich2018}
{Pillepich}, A., {Springel}, V., {Nelson}, D., {et~al.} 2018,
  \href{http://dx.doi.org/10.1093/mnras/stx2656}{\JournalTitle{\mnras}, 473,
  4077}

\bibitem[{{Pillepich} {et~al.}(2019){Pillepich}, {Nelson}, {Springel},
  {Pakmor}, {Torrey}, {Weinberger}, {Vogelsberger}, {Marinacci}, {Genel}, {van
  der Wel}, \& {Hernquist}}]{pillepich2019}
{Pillepich}, A., {Nelson}, D., {Springel}, V., {et~al.} 2019,
  \href{http://dx.doi.org/10.1093/mnras/stz2338}{\JournalTitle{\mnras}, 490,
  3196}

\bibitem[{{Planck Collaboration} {et~al.}(2020){Planck Collaboration},
  {Aghanim}, {Akrami}, {Ashdown}, {Aumont}, {Baccigalupi}, {Ballardini},
  {Banday}, {Barreiro}, {Bartolo}, {Basak}, {Battye}, {Benabed}, {Bernard},
  {Bersanelli}, {Bielewicz}, {Bock}, {Bond}, {Borrill}, {Bouchet}, {Boulanger},
  {Bucher}, {Burigana}, {Butler}, {Calabrese}, {Cardoso}, {Carron},
  {Challinor}, {Chiang}, {Chluba}, {Colombo}, {Combet}, {Contreras}, {Crill},
  {Cuttaia}, {de Bernardis}, {de Zotti}, {Delabrouille}, {Delouis}, {Di
  Valentino}, {Diego}, {Dor{\'e}}, {Douspis}, {Ducout}, {Dupac}, {Dusini},
  {Efstathiou}, {Elsner}, {En{\ss}lin}, {Eriksen}, {Fantaye}, {Farhang},
  {Fergusson}, {Fernandez-Cobos}, {Finelli}, {Forastieri}, {Frailis},
  {Fraisse}, {Franceschi}, {Frolov}, {Galeotta}, {Galli}, {Ganga},
  {G{\'e}nova-Santos}, {Gerbino}, {Ghosh}, {Gonz{\'a}lez-Nuevo}, {G{\'o}rski},
  {Gratton}, {Gruppuso}, {Gudmundsson}, {Hamann}, {Handley}, {Hansen},
  {Herranz}, {Hildebrandt}, {Hivon}, {Huang}, {Jaffe}, {Jones}, {Karakci},
  {Keih{\"a}nen}, {Keskitalo}, {Kiiveri}, {Kim}, {Kisner}, {Knox},
  {Krachmalnicoff}, {Kunz}, {Kurki-Suonio}, {Lagache}, {Lamarre}, {Lasenby},
  {Lattanzi}, {Lawrence}, {Le Jeune}, {Lemos}, {Lesgourgues}, {Levrier},
  {Lewis}, {Liguori}, {Lilje}, {Lilley}, {Lindholm}, {L{\'o}pez-Caniego},
  {Lubin}, {Ma}, {Mac{\'\i}as-P{\'e}rez}, {Maggio}, {Maino}, {Mandolesi},
  {Mangilli}, {Marcos-Caballero}, {Maris}, {Martin}, {Martinelli},
  {Mart{\'\i}nez-Gonz{\'a}lez}, {Matarrese}, {Mauri}, {McEwen}, {Meinhold},
  {Melchiorri}, {Mennella}, {Migliaccio}, {Millea}, {Mitra},
  {Miville-Desch{\^e}nes}, {Molinari}, {Montier}, {Morgante}, {Moss}, {Natoli},
  {N{\o}rgaard-Nielsen}, {Pagano}, {Paoletti}, {Partridge}, {Patanchon},
  {Peiris}, {Perrotta}, {Pettorino}, {Piacentini}, {Polastri}, {Polenta},
  {Puget}, {Rachen}, {Reinecke}, {Remazeilles}, {Renzi}, {Rocha}, {Rosset},
  {Roudier}, {Rubi{\~n}o-Mart{\'\i}n}, {Ruiz-Granados}, {Salvati}, {Sandri},
  {Savelainen}, {Scott}, {Shellard}, {Sirignano}, {Sirri}, {Spencer},
  {Sunyaev}, {Suur-Uski}, {Tauber}, {Tavagnacco}, {Tenti}, {Toffolatti},
  {Tomasi}, {Trombetti}, {Valenziano}, {Valiviita}, {Van Tent}, {Vibert},
  {Vielva}, {Villa}, {Vittorio}, {Wandelt}, {Wehus}, {White}, {White},
  {Zacchei}, \& {Zonca}}]{planck2020}
{Planck Collaboration}, {Aghanim}, N., {Akrami}, Y., {et~al.} 2020,
  \href{http://dx.doi.org/10.1051/0004-6361/201833910}{\JournalTitle{\aap},
  641, A6}

\bibitem[{{Pontzen} \& {Governato}(2012)}]{pontzen2012}
{Pontzen}, A., \& {Governato}, F. 2012,
  \href{http://dx.doi.org/10.1111/j.1365-2966.2012.20571.x}{\JournalTitle{\mnras},
  421, 3464}

\bibitem[{{Price} \& {Monaghan}(2007)}]{Price2007}
{Price}, D.~J., \& {Monaghan}, J.~J. 2007,
  \href{http://dx.doi.org/10.1111/j.1365-2966.2006.11241.x}{\JournalTitle{\mnras},
  374, 1347}

\bibitem[{{Pucha} {et~al.}(2019){Pucha}, {Carlin}, {Willman}, {Strader},
  {Sand}, {Bechtol}, {Brodie}, {Crnojevi{\'c}}, {Forbes}, {Garling}, {Hargis},
  {Peter}, \& {Romanowsky}}]{pucha2019}
{Pucha}, R., {Carlin}, J.~L., {Willman}, B., {et~al.} 2019,
  \href{http://dx.doi.org/10.3847/1538-4357/ab29fb}{\JournalTitle{\apj}, 880,
  104}

\bibitem[{{Purcell} {et~al.}(2007){Purcell}, {Bullock}, \&
  {Zentner}}]{purcell2007}
{Purcell}, C.~W., {Bullock}, J.~S., \& {Zentner}, A.~R. 2007,
  \href{http://dx.doi.org/10.1086/519787}{\JournalTitle{\apj}, 666, 20}

\bibitem[{{Read} \& {Gilmore}(2005)}]{read2005}
{Read}, J.~I., \& {Gilmore}, G. 2005,
  \href{http://dx.doi.org/10.1111/j.1365-2966.2004.08424.x}{\JournalTitle{\mnras},
  356, 107}

\bibitem[{{Read} {et~al.}(2006){Read}, {Pontzen}, \& {Viel}}]{read2006}
{Read}, J.~I., {Pontzen}, A.~P., \& {Viel}, M. 2006,
  \href{http://dx.doi.org/10.1111/j.1365-2966.2006.10720.x}{\JournalTitle{\mnras},
  371, 885}

\bibitem[{{Rong} {et~al.}(2019){Rong}, {Dong}, {Puzia}, {Galaz}, {Cao}, {Van
  Der Burg}, {Sifon}, {Mancera Pina}, {Marcelo}, {D'ago}, {Zhang}, {Johnston},
  \& {Eigenthaler}}]{rong2019}
{Rong}, Y., {Dong}, X.-Y., {Puzia}, T.~H., {et~al.} 2019, \JournalTitle{arXiv
  e-prints}, arXiv:1907.10079

\bibitem[{{Roychowdhury} {et~al.}(2013){Roychowdhury}, {Chengalur},
  {Karachentsev}, \& {Kaisina}}]{roychowdhury2013}
{Roychowdhury}, S., {Chengalur}, J.~N., {Karachentsev}, I.~D., \& {Kaisina},
  E.~I. 2013,
  \href{http://dx.doi.org/10.1093/mnrasl/slt123}{\JournalTitle{\mnras}, 436,
  L104}

\bibitem[{{S{\'a}nchez-Janssen} {et~al.}(2010){S{\'a}nchez-Janssen},
  {M{\'e}ndez-Abreu}, \& {Aguerri}}]{sanchezjanssen2010}
{S{\'a}nchez-Janssen}, R., {M{\'e}ndez-Abreu}, J., \& {Aguerri}, J.~A.~L. 2010,
  \href{http://dx.doi.org/10.1111/j.1745-3933.2010.00883.x}{\JournalTitle{\mnras},
  406, L65}

\bibitem[{{Sanderson} {et~al.}(2018){Sanderson}, {Garrison-Kimmel}, {Wetzel},
  {Keung Chan}, {Hopkins}, {Kere{\v{s}}}, {Escala}, {Faucher-Gigu{\`e}re}, \&
  {Ma}}]{sanderson2018}
{Sanderson}, R.~E., {Garrison-Kimmel}, S., {Wetzel}, A., {et~al.} 2018,
  \href{http://dx.doi.org/10.3847/1538-4357/aaeb33}{\JournalTitle{\apj}, 869,
  12}

\bibitem[{{Sanderson} {et~al.}(2020){Sanderson}, {Wetzel}, {Loebman}, {Sharma},
  {Hopkins}, {Garrison-Kimmel}, {Faucher-Gigu{\`e}re}, {Kere{\v{s}}}, \&
  {Quataert}}]{sanderson2020}
{Sanderson}, R.~E., {Wetzel}, A., {Loebman}, S., {et~al.} 2020,
  \href{http://dx.doi.org/10.3847/1538-4365/ab5b9d}{\JournalTitle{\apjs}, 246,
  6}

\bibitem[{{Simonneau} {et~al.}(1998){Simonneau}, {Varela}, \&
  {Munoz-Tunon}}]{simonneau1998}
{Simonneau}, E., {Varela}, A.~M., \& {Munoz-Tunon}, C. 1998,
  \JournalTitle{Nuovo Cimento B Serie}, 113B, 927

\bibitem[{{Smith} {et~al.}(2020){Smith}, {Bryan}, {Somerville}, {Hu},
  {Teyssier}, {Burkhart}, \& {Hernquist}}]{smith2020}
{Smith}, M.~C., {Bryan}, G.~L., {Somerville}, R.~S., {et~al.} 2020,
  \JournalTitle{arXiv e-prints}, arXiv:2009.11309

\bibitem[{{Smith} {et~al.}(2019){Smith}, {Sijacki}, \& {Shen}}]{smith2019}
{Smith}, M.~C., {Sijacki}, D., \& {Shen}, S. 2019,
  \href{http://dx.doi.org/10.1093/mnras/stz599}{\JournalTitle{\mnras}, 485,
  3317}

\bibitem[{{Sparre} {et~al.}(2017){Sparre}, {Hayward}, {Feldmann},
  {Faucher-Gigu{\`e}re}, {Muratov}, {Kere{\v{s}}}, \& {Hopkins}}]{sparre2017}
{Sparre}, M., {Hayward}, C.~C., {Feldmann}, R., {et~al.} 2017,
  \href{http://dx.doi.org/10.1093/mnras/stw3011}{\JournalTitle{\mnras}, 466,
  88}

\bibitem[{{Springel}(2005)}]{Springel2005gadget2}
{Springel}, V. 2005,
  \href{http://dx.doi.org/10.1111/j.1365-2966.2005.09655.x}{\JournalTitle{\mnras},
  364, 1105}

\bibitem[{{Starkenburg} \& {Helmi}(2015)}]{starkenburg2015}
{Starkenburg}, T.~K., \& {Helmi}, A. 2015,
  \href{http://dx.doi.org/10.1051/0004-6361/201425082}{\JournalTitle{\aap},
  575, A59}

\bibitem[{{Starkenburg} {et~al.}(2016){Starkenburg}, {Helmi}, \&
  {Sales}}]{starkenburg2016}
{Starkenburg}, T.~K., {Helmi}, A., \& {Sales}, L.~V. 2016,
  \href{http://dx.doi.org/10.1051/0004-6361/201527247}{\JournalTitle{\aap},
  587, A24}

\bibitem[{{Stinson} {et~al.}(2009){Stinson}, {Dalcanton}, {Quinn}, {Gogarten},
  {Kaufmann}, \& {Wadsley}}]{stinson2009}
{Stinson}, G.~S., {Dalcanton}, J.~J., {Quinn}, T., {et~al.} 2009,
  \href{http://dx.doi.org/10.1111/j.1365-2966.2009.14555.x}{\JournalTitle{\mnras},
  395, 1455}

\bibitem[{{Strader} {et~al.}(2012){Strader}, {Seth}, \&
  {Caldwell}}]{strader2012}
{Strader}, J., {Seth}, A.~C., \& {Caldwell}, N. 2012,
  \href{http://dx.doi.org/10.1088/0004-6256/143/2/52}{\JournalTitle{\aj}, 143,
  52}

\bibitem[{{Su} {et~al.}(2017){Su}, {Hopkins}, {Hayward}, {Faucher-Gigu{\`e}re},
  {Kere{\v{s}}}, {Ma}, \& {Robles}}]{su2017}
{Su}, K.-Y., {Hopkins}, P.~F., {Hayward}, C.~C., {et~al.} 2017,
  \href{http://dx.doi.org/10.1093/mnras/stx1463}{\JournalTitle{\mnras}, 471,
  144}

\bibitem[{Van Der~Walt {et~al.}(2011)Van Der~Walt, Colbert, \&
  Varoquaux}]{van2011numpy}
Van Der~Walt, S., Colbert, S.~C., \& Varoquaux, G. 2011,
  \JournalTitle{Computing in Science \& Engineering}, 13, 22

\bibitem[{{van der Wel} {et~al.}(2014){van der Wel}, {Chang}, {Bell}, {Holden},
  {Ferguson}, {Giavalisco}, {Rix}, {Skelton}, {Whitaker}, {Momcheva},
  {Brammer}, {Kassin}, {Martig}, {Dekel}, {Ceverino}, {Koo}, {Mozena}, {van
  Dokkum}, {Franx}, {Faber}, \& {Primack}}]{vanderwel2014}
{van der Wel}, A., {Chang}, Y.-Y., {Bell}, E.~F., {et~al.} 2014,
  \href{http://dx.doi.org/10.1088/2041-8205/792/1/L6}{\JournalTitle{\apjl},
  792, L6}

\bibitem[{{Vansevi{\v{c}}ius} {et~al.}(2004){Vansevi{\v{c}}ius}, {Arimoto},
  {Hasegawa}, {Ikuta}, {Jablonka}, {Narbutis}, {Ohta}, {Stonkut{\.{e}}},
  {Tamura}, {Vansevi{\v{c}}ius}, \& {Yamada}}]{vansevicius2004}
{Vansevi{\v{c}}ius}, V., {Arimoto}, N., {Hasegawa}, T., {et~al.} 2004,
  \href{http://dx.doi.org/10.1086/423802}{\JournalTitle{\apjl}, 611, L93}

\bibitem[{{Weisz} \& {Boylan-Kolchin}(2017)}]{2017MNRAS.469L..83W}
{Weisz}, D.~R., \& {Boylan-Kolchin}, M. 2017,
  \href{http://dx.doi.org/10.1093/mnrasl/slx043}{\JournalTitle{\mnras}, 469,
  L83}

\bibitem[{{Wetzel} \& {Garrison-Kimmel}(2020)}]{wetzel2020}
{Wetzel}, A., \& {Garrison-Kimmel}, S. 2020, {GizmoAnalysis: Read and analyze
  Gizmo simulations}

\bibitem[{{Wetzel} {et~al.}(2016){Wetzel}, {Hopkins}, {Kim},
  {Faucher-Gigu{\`e}re}, {Kere{\v{s}}}, \& {Quataert}}]{wetzel2016}
{Wetzel}, A.~R., {Hopkins}, P.~F., {Kim}, J.-h., {et~al.} 2016,
  \href{http://dx.doi.org/10.3847/2041-8205/827/2/L23}{\JournalTitle{\apjl},
  827, L23}

\bibitem[{{Wheeler} {et~al.}(2017){Wheeler}, {Pace}, {Bullock},
  {Boylan-Kolchin}, {O{\~n}orbe}, {Elbert}, {Fitts}, {Hopkins}, \&
  {Kere{\v{s}}}}]{wheeler2017}
{Wheeler}, C., {Pace}, A.~B., {Bullock}, J.~S., {et~al.} 2017,
  \href{http://dx.doi.org/10.1093/mnras/stw2583}{\JournalTitle{\mnras}, 465,
  2420}

\bibitem[{{Wheeler} {et~al.}(2019){Wheeler}, {Hopkins}, {Pace},
  {Garrison-Kimmel}, {Boylan-Kolchin}, {Wetzel}, {Bullock}, {Kere{\v{s}}},
  {Faucher-Gigu{\`e}re}, \& {Quataert}}]{wheeler2019}
{Wheeler}, C., {Hopkins}, P.~F., {Pace}, A.~B., {et~al.} 2019,
  \href{http://dx.doi.org/10.1093/mnras/stz2887}{\JournalTitle{\mnras}, 490,
  4447}

\bibitem[{{Wright} {et~al.}(2020){Wright}, {Tremmel}, {Brooks}, {Munshi},
  {Nagai}, {Sharma}, \& {Quinn}}]{wright2020}
{Wright}, A.~C., {Tremmel}, M., {Brooks}, A.~M., {et~al.} 2020,
  \JournalTitle{arXiv e-prints}, arXiv:2005.07634

\bibitem[{{Yoachim} \& {Dalcanton}(2006)}]{yoachim2006}
{Yoachim}, P., \& {Dalcanton}, J.~J. 2006,
  \href{http://dx.doi.org/10.1086/497970}{\JournalTitle{\aj}, 131, 226}

\bibitem[{{Yu} {et~al.}(2021){Yu}, {Bullock}, {Klein}, {Stern}, {Wetzel}, {Ma},
  {Moreno}, {Hafen}, {Gurvich}, {Hopkins}, {Kere{\v{s}}},
  {Faucher-Gigu{\`e}re}, {Feldmann}, \& {Quataert}}]{yu2021}
{Yu}, S., {Bullock}, J.~S., {Klein}, C., {et~al.} 2021, \JournalTitle{arXiv
  e-prints}, arXiv:2103.03888

\bibitem[{{Zaritsky} {et~al.}(2000){Zaritsky}, {Harris}, {Grebel}, \&
  {Thompson}}]{zaritsky2000}
{Zaritsky}, D., {Harris}, J., {Grebel}, E.~K., \& {Thompson}, I.~B. 2000,
  \href{http://dx.doi.org/10.1086/312649}{\JournalTitle{\apjl}, 534, L53}

\bibitem[{{Zhang} {et~al.}(2019){Zhang}, {Primack}, {Faber}, {Koo}, {Dekel},
  {Chen}, {Ceverino}, {Chang}, {Fang}, {Guo}, {Lin}, \& {Wel}}]{zhang2019}
{Zhang}, H., {Primack}, J.~R., {Faber}, S.~M., {et~al.} 2019,
  \href{http://dx.doi.org/10.1093/mnras/stz339}{\JournalTitle{\mnras}, 484,
  5170}

\end{thebibliography}

\appendix

\section{Placing Limits on the Presence of a Non-Disky Minority Population}\label{s:toymodel}
In the FIRE\rrrthree{-2} simulations, we find two distinct morphological classes of dwarf galaxies -- those 
that are able to form young stellar disks and a round old stellar halo, and those that are
characterized by highly \ndd{} shapes at all ages. In the observational analysis of
\cite{kadofong2020}, the intrinsic shape population was modeled as a single multivariate Gaussian.
It is thus of interest to make an estimate of the maximum fraction of \ndd{} galaxies that
could be present in a dwarf sample without affecting the inferred
observational intrinsic shape distribution.
Here, we present mock trials to argue that no more than 10\% of a 
\ndd{} population drawn from the FIRE\rrrthree{-2} \ndd{} dwarfs 
could exist in a disky population dominated by the FIRE\rrrthree{-2} disky
dwarfs without changing the results of the observational 
inference method used in \cite{kadofong2020}.

To do so, we first briefly summarize the observational inference method; 
a full explanation can be found in Section 4 of \cite{kadofong2020}. As was demonstrated by 
\cite{simonneau1998}, it can be shown that the projected axis ratio $\rm b/a$ of an ellipsoid
is an analytic function of its intrinsic axes ratios, $\rm B/A$ and $\rm C/A$, and the 
observer's viewing angle $(\theta,\phi)$. A given intrinsic axis ratio distribution can then be
quickly turned into a distribution of projected ellipticities, assuming that the viewing angle is
distributed uniformly over a sphere. This ellipticity distribution can then be used, in conjunction
with adopted flat priors over the relevant shape parameters, to infer the distribution of 
intrinsic shapes for a galaxy sample given a chosen functional form for that distribution via
Markov Chain Monte Carlo sampling \citep[MCMC via \textsf{emcee},][]{foremanmackey2013}. 
\cite{kadofong2020} adopted a multivariate Gaussian for the inference with parameters 
$\vec \alpha = \{\rm \mu_B, \mu_C, \sigma_B,\sigma_C\}$, where $\rm \mu_X$ and $\rm \sigma_X$ are
the mean and standard deviation of the Gaussian characterizing the
axis ratio $\rm X/A$ distribution, respectively. This choice enforces a 
singly-peaked distribution in intrinsic shape space.

\begin{figure*}[htb]
\centering     
\includegraphics[width=\linewidth]{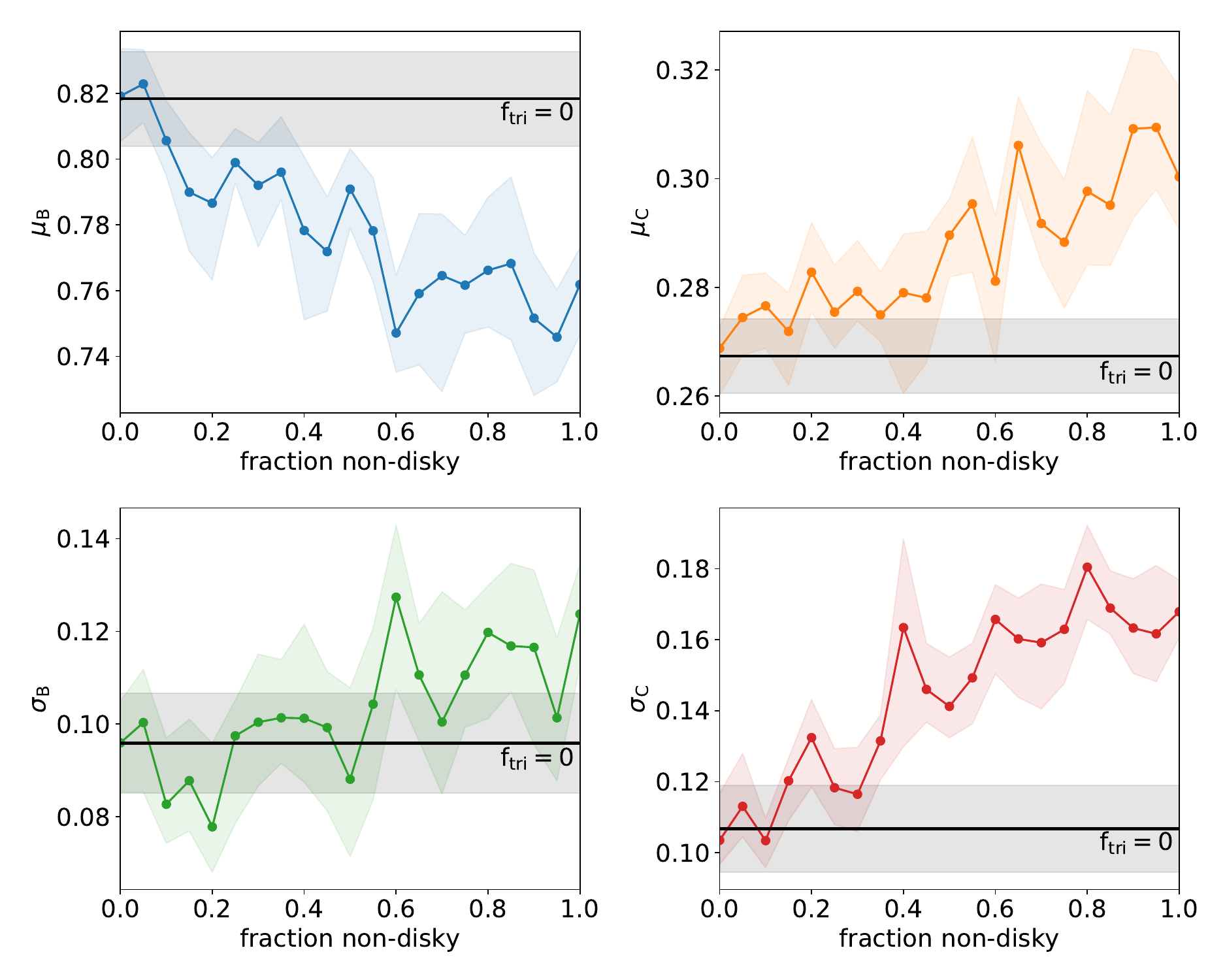}
\caption{ 
    The maximum a posteriori (solid curves) and standard deviation (shaded regions) estimates of
    the observational intrinsic shape parameters as a function of the fraction of the mock sample
    that is drawn from the \ndd{} galaxy distribution. The solid black line (region) in each
    panel shows the maximum a posteriori estimate (standard deviation) of the $\rm f_{\rm \ndd{}}=0$ sample. Clockwise from top left, we show: $\rm \mu_B$, the mean axis ratio $\rm B/A$, $\rm\mu_C$, the mean axis ratio $\rm C/A$, $\rm \sigma_B$, the standard deviation of B/A, and $\rm \sigma_C$, the standard deviation of C/A. As the \ndd{} fraction increases, the mean
    B/A decreases and mean C/A increases, consistent with a more \ndd{} shape overall.
    }
\label{f:chainstats}
\end{figure*}

To investigate the impact of a minority population of \ndd{} galaxies, we invoke the following toy
model. First, to characterize our two underlying intrinsic shape distributions, we measure the 
mean and covariance matrices of the \ndd{} and disky FIRE\rrrthree{-2} dwarfs separately. Then, we generate 
a sample from those shape distributions with a given fraction of \ndd{} galaxies ranging between
0 and 1, and infer the intrinsic shapes of the resulting ellipticity distribution using the same
framework as \cite{kadofong2020}. 
The mock samples are of size N=4600, selected to be approximately equal to the largest
mass bin of \cite{kadofong2020}. \autoref{f:chainstats} shows the evolution of the inferred 
shape parameters as a function of the \ndd{} fraction in the mock sample. As expected, we find 
that the mean B/A decreases and mean C/A increases as the \ndd{} fraction increases. 
Interestingly, although the inferred mean shapes of the pure disk population are in good agreement
with the true values of the FIRE\rrrthree{-2} sample, the inferred mean shapes of the pure \ndd{} population
are somewhat less \ndd{} than the true values of the FIRE\rrrthree{-2} sample. This is likely due to a
significant covariance between B/A and C/A in the \ndd{} sample that is not captured by the
model.

\begin{figure*}[htb]
\centering     
\includegraphics[width=0.5\linewidth]{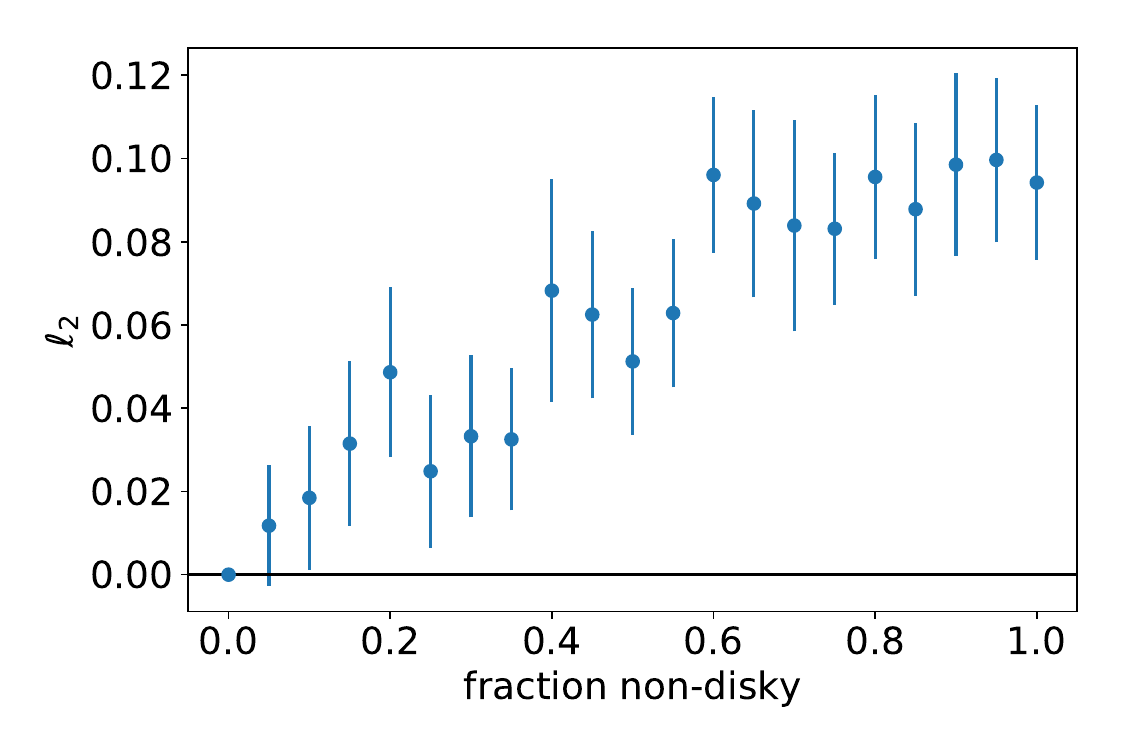}
\caption{ 
    The L2-norm, $\ell_2$, calculated between the inferred shape parameters of the mock 
    observational sample as a function of the mock sample that is drawn from the \ndd{} 
    galaxy distribution. The errors reflect the propagation of errors using the standard deviation
    of the MCMC chains as the uncertainty on the inferred shape parameters.
    }
\label{f:l2norm}
\end{figure*}

To quantify the \ndd{} fraction at which the mock observational inference changes significantly
from the results for a pure disk ($\rm f_{\ndd{}}=0$) sample, we compute the $\ell_2$ norm 
between the mean values of each sample and the pure disk sample. The uncertainty $\sigma_{\ell_2}$
is the propagation of errors where we adopt the standard deviation of the MCMC chains as the 
uncertainty on the estimates of the shape parameters. We show the results of this analysis
in \autoref{f:l2norm} -- we find that the inferred parameters change significantly when the
\ndd{} fraction exceeds $\rm f_{\ndd{}} =0.1$. Thus, though it is possible for a 
sub-population of morphologically distinct galaxies to exist undetected within a generally disky
population, this possible fraction is much smaller than the realized $\sim60\%$ \ndd{} fraction
seen in the FIRE\rrrthree{-2} galaxies.

\section{\rrrtwo{A comparison between direct cumulative and inferred differential shape measurements}}\label{s:appendix:inference}
\rrrtwo{As discussed in the main text of this work, the methodology that we use to measure
the intrinsic shapes of the FIRE\rrrthree{-2} galaxies differs significantly from the methodology
used to infer the shape distribution of the observed dwarf sample in
\cite{kadofong2020} (see \autoref{s:methods:intrinsicshape}). The observational 
methodology is fundamentally unsuitable for an analysis of the FIRE-2 dwarfs
for three main reasons. First, our sample includes only nine independent initial 
conditions (i.e. not considering runs with different physics as separate), which is
not sufficient to fully sample the underlying distribution of three-dimensional shapes.
This runs counter to the assumption of a well-sampled population 
made in the observational inversion technique
of \cite{kadofong2020}; even though the simulated galaxies may be projected at 
an unlimited number of viewing angles to generate a smooth distribution in ellipticity, 
the joint ellipticity distribution of the projected FIRE-2 dwarfs would still be generated
from a very low number of unique three-dimensional shapes. Second, as mentioned in the
main text, we choose to use cumulative shapes to prevent instability in the shape 
measurements due to a low number of star particles in the outskirts \citep[see 
also][]{allgood2006}. Because this inference method adds no new additional information,
it is likely to suffer from the same instability. Finally, the simulations offer a 
more direct method to compute the three-dimensional galaxy shapes; using the 
observational inference method will work to add noise to our measurements, but, as
we have detailed above, not contribute significant insight.}

\begin{figure*}[htb]
\centering     
\includegraphics[width=.9\linewidth]{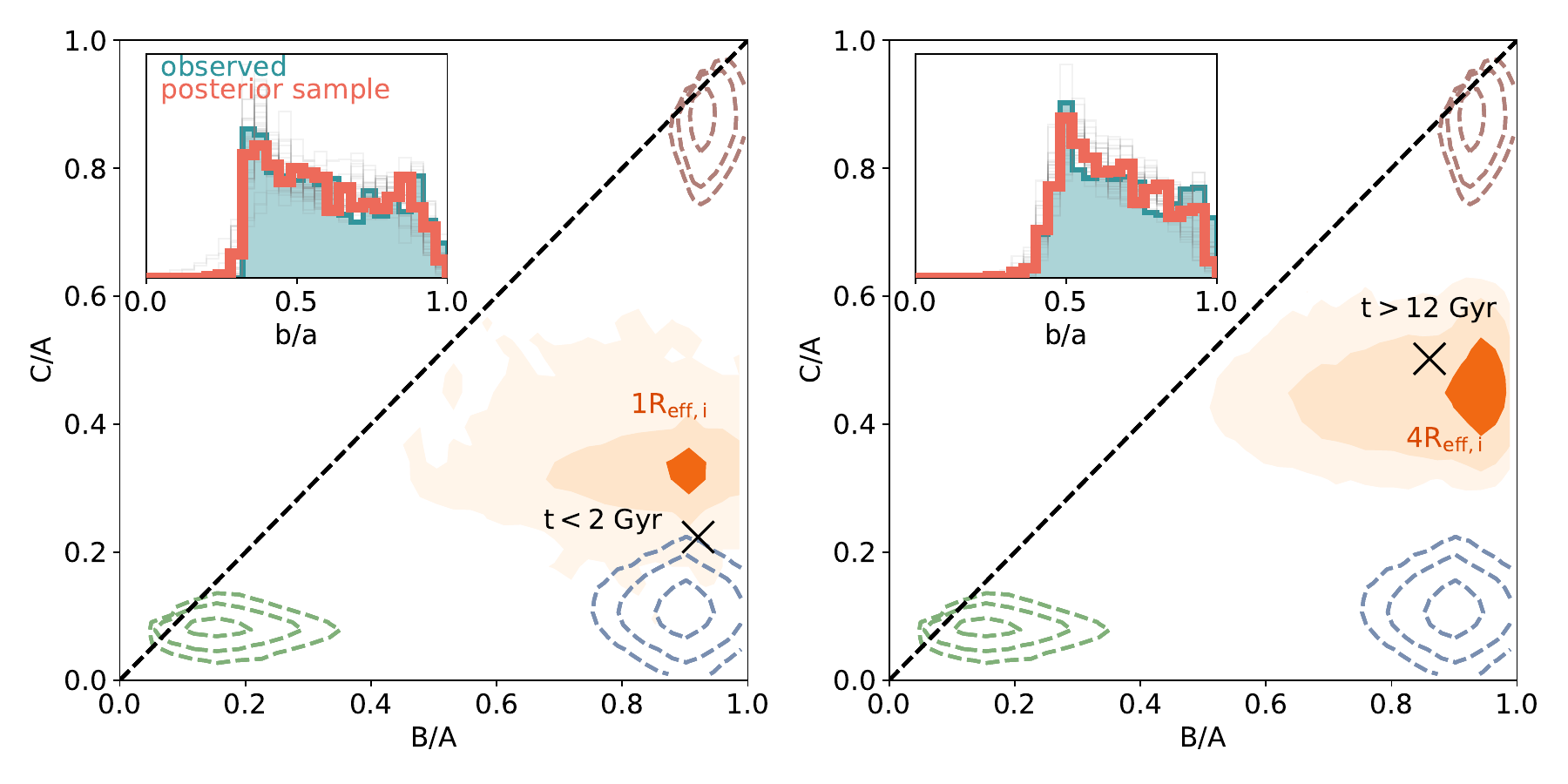}
\caption{ 
    \rrrtwo{\textit{Left:} the results of the mock observational inference for m11h (CR+) at 
    one effective radius. The inset panel shows the distribution of the projected axis
    ratio b/a for 1000 instances of random projections of the FIRE-2 dwarf in green; the
    orange curve shows the model output from the inference machinery (600 steps with 300 burn-in; walker convergence is checked manually). Light grey curves
    show individual draws from the posterior. The main panel shows the same model in
    intrinsic axis ratio space; the direct cumulative measurement for the 
    young (age $< 2$ Gyr) star particles is shown by the black X. The colored, dashed contours
    show results for toy models of a disk (blue), spheroid (red) and prolate (green) galaxy
    under HSC wide imaging conditions. \textit{Right:} the same, but for differential
    measurements at four effective radii and cumulative measurements of old (age $> 12$ Gyr)
    star particles. In both cases, the differential and cumulative shape measurements 
    well-trace each other. }
    }
\label{f:appendix_inversion}
\end{figure*}

\rrrtwo{Nevertheless, we believe that it is informative to the reader to show that the
 age-separated, cumulative shape measurements for a 
single galaxy trace the radial differential measurements one would have derived for
that same single galaxy. To do so, we compute 1000 images in the HSC i-band 
of the relatively massive and disky dwarf m11h (CR+) at a random set of viewing angles
distributed isotropically over the unit sphere. We then use the same isophotal fitting
technique of \cite{kadofong2020} to compute the ellipticity distribution of these projected
images from one to four effective radii. We show the results of this test in
\autoref{f:appendix_inversion}. We find that the age separation approach well-traces 
the results of the radial inference, though we again emphasize that these approaches
are not expected to yield identical results. We thus confirm in a more direct way that the
measurements from the age-separated cumulative shapes can be meaningfully compared to the
differential shape measurements from \cite{kadofong2020}.
}

\section{Projection Gallery of Disky and non-Disky Dwarfs}
\begin{figure*}[htb]
\centering     
\includegraphics[width=\linewidth]{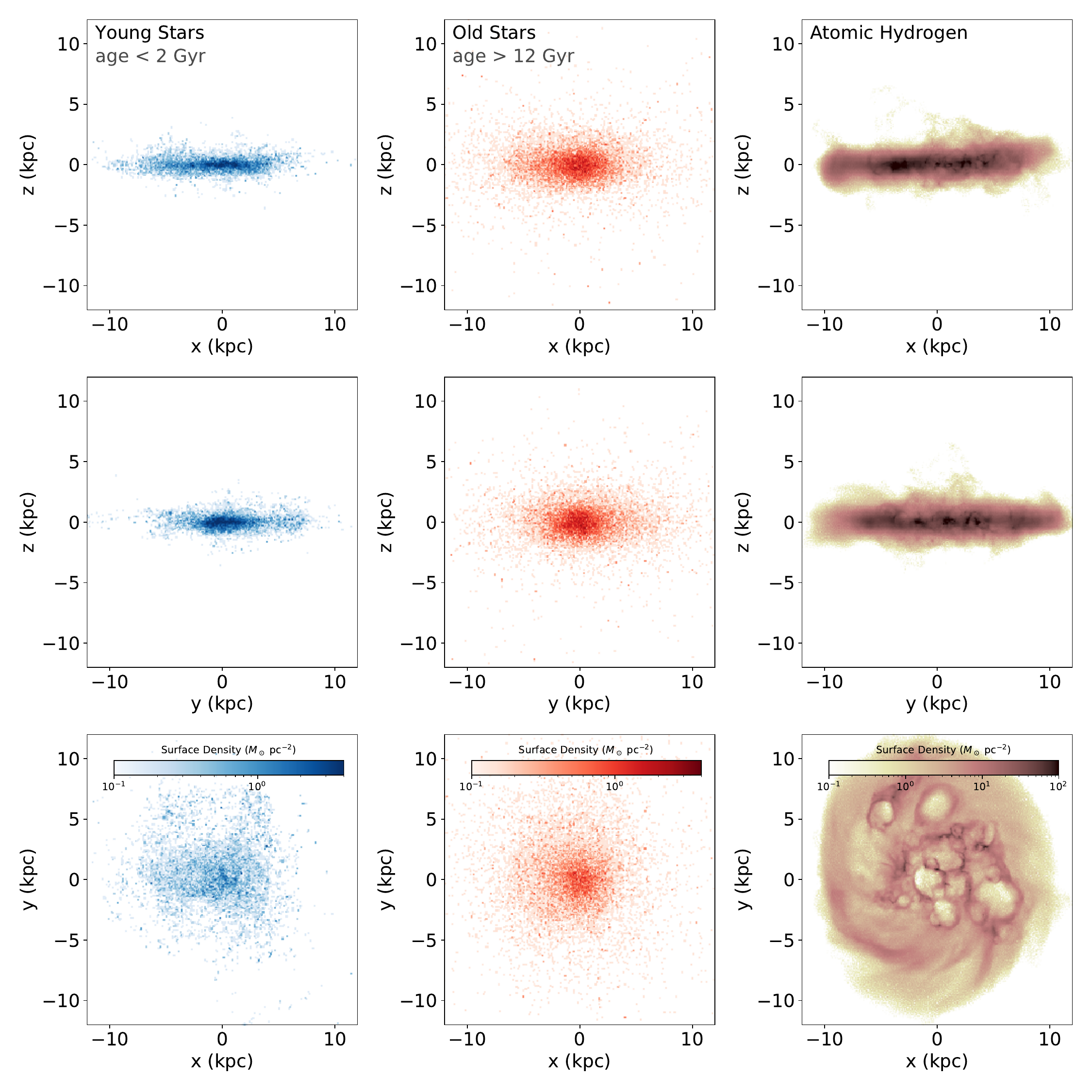}
\caption{ 
    Projections of young stars (left), old stars (middle), and HI 
    gas (right) for the disky dwarf m11b (Hydro+, no MD) in the
    same format as \autoref{f:projections_m11h}. 
    }
\label{f:projections_m11b}
\end{figure*}

\begin{figure*}[htb]
\centering     
\includegraphics[width=\linewidth]{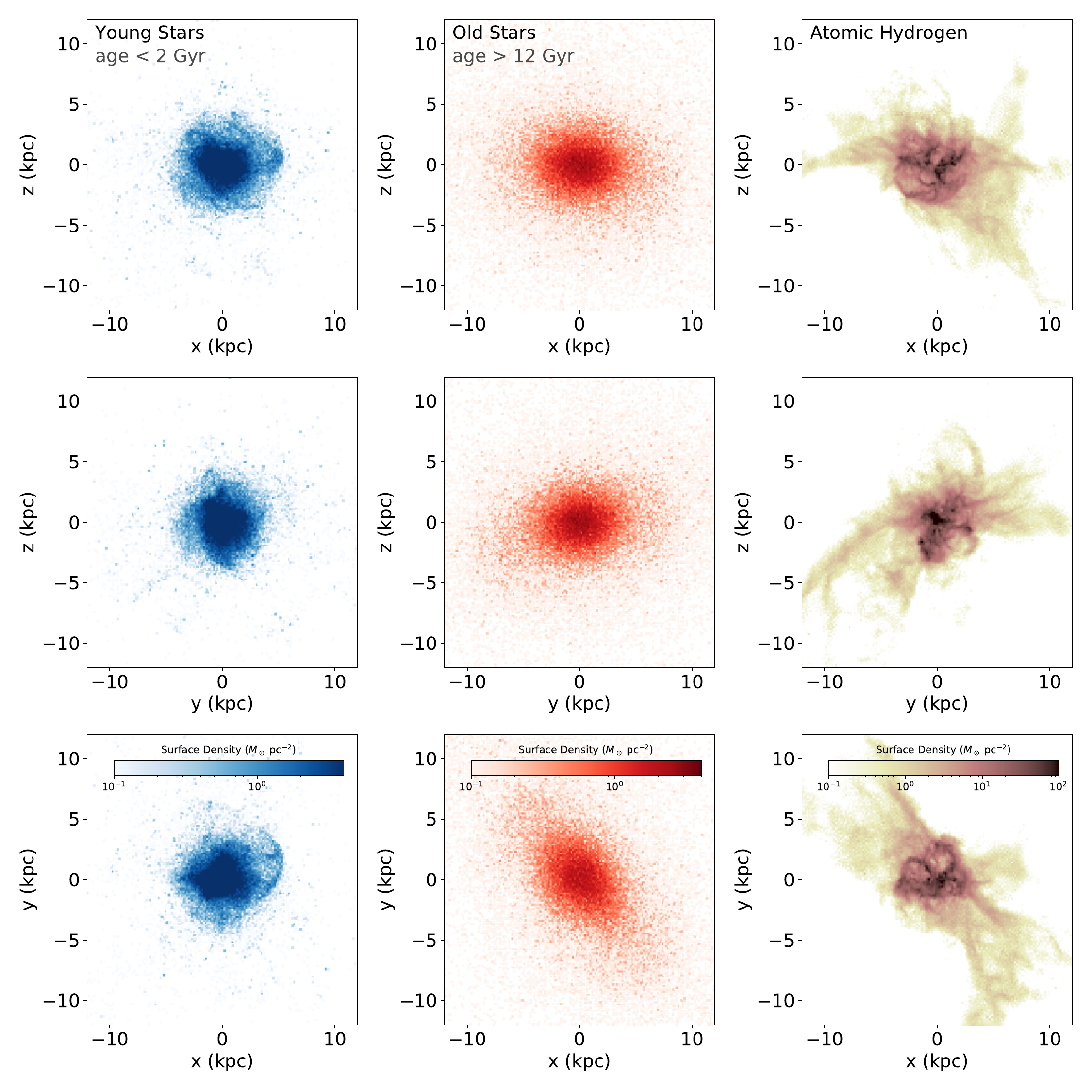}
\caption{ 
    Projections of young stars (left), old stars (middle), and HI 
    gas (right) for the \ndd{} dwarf m11c (Hydro+, no MD) in the
    same format as \autoref{f:projections_m11h}.  
    }
\label{f:projections_m11c}
\end{figure*}

To give the reader a wider sense of the morphologies spanned by the
FIRE\rrrthree{-2} dwarfs, we show the $z=0$ projections of young stars, old stars, and
HI in the same format as \autoref{f:projections_m11h}
for the disky dwarf m11b (Hydro+, no MD) in \autoref{f:projections_m11b} 
and the \ndd{} dwarf m11c (Hydro+, no MD) in \autoref{f:projections_m11c}.

\end{document}